\documentclass[lettersize,journal]{IEEEtran}
\usepackage{enumitem}
\usepackage{amsfonts}
\usepackage{enumitem,kantlipsum}
\usepackage{textcomp}
\usepackage{stfloats}
\usepackage{url}
\usepackage{caption}
\usepackage{graphics}
\usepackage{tabularx}
\usepackage[dvipsnames,table]{xcolor}
\usepackage{verbatim}
\usepackage{pifont}
\usepackage{float}
\usepackage{amsmath}
\usepackage{cancel}
\usepackage{libertinust1math}
\usepackage{pgfplots}
\usepackage{amsmath}
\usepackage{forest}
\usepackage{graphicx}
\usepackage{cite}
\usepackage{etoolbox}
\usepackage{graphicx}
\usepackage{float}
\usepackage{lscape}
\usepackage{longtable}
\usepackage{threeparttable}
\usepackage{ltxtable}
\usepackage{hhline}
\usepackage{caption}
\usepackage{wasysym}
\usepackage{subcaption}
\usepackage{algorithm} 
\usepackage{color} 
\usetikzlibrary{patterns}
\usepackage{colortbl}
\usepackage{algpseudocode}
\usepackage{verbatim}
\usepackage{lipsum}
\usetikzlibrary{automata, arrows.meta, positioning}
\usetikzlibrary{decorations.pathreplacing}
\usepackage{amsthm}
\usepackage{booktabs}
\usepackage{amsmath}
\usepackage{algorithm}
\usepackage{inputenc}
\usepackage{adjustbox}
\usepackage{url} 
\usepackage{multirow}
\usepackage{color,soul}
\usepackage{lipsum}
\usepackage{mathtools}
\usepackage{cuted}
\usepackage{xcolor}
\usepackage{wasysym}
\usepackage{pgfplotstable}
\usepackage[normalem]{ulem}
\usetikzlibrary{arrows.meta}
\usepackage{smartdiagram}
\usepackage{tikz}
\usetikzlibrary{mindmap, shapes, shadows}
\usepackage{xcolor}

\usepackage{enumitem}
\usepackage{amsfonts}
\usepackage{enumitem,kantlipsum}
\usepackage{textcomp}
\usepackage{stfloats}
\usepackage{url}
\usepackage{caption}
\usepackage{graphics}
\usepackage{tabularx}
\usepackage[dvipsnames,table]{xcolor}
\usepackage{verbatim}
\usepackage{pifont}
\usepackage{float}
\usepackage{amsmath}
\usepackage{cancel}
\usepackage{libertinust1math}
\usepackage{pgfplots}
\usepackage{amsmath}
\usepackage{forest}
\usepackage{graphicx}
\usepackage{cite}
\usepackage{etoolbox}
\usepackage{graphicx}
\usepackage{float}
\usepackage{lscape}
\usepackage{longtable}
\usepackage{ltxtable}
\usepackage{hhline}
\usepackage{caption}
\usepackage{wasysym}
\usepackage{subcaption}
\usepackage{algorithm} 
\usepackage{color} 
\usepackage{colortbl}
\usepackage{algpseudocode}
\usepackage{verbatim}
\usepackage{lipsum}
\usetikzlibrary{automata, arrows.meta, positioning}
\usetikzlibrary{decorations.pathreplacing}
\usepackage{amsthm}
\usepackage{booktabs}
\usepackage{amsmath}
\usepackage{algorithm}
\usepackage{inputenc}
\usepackage{adjustbox}
\usepackage{url} 
\usepackage{multirow}
\usepackage{color,soul}
\usepackage{lipsum}
\usepackage{mathtools}
\usepackage{cuted}
\usepackage{xcolor}
\usepackage{wasysym}
\usepackage{tikz}
\usepackage{adjustbox}
\usetikzlibrary{shapes.geometric, arrows}

\usepackage{url}    
\usepackage{xurl}   

\usepackage{tikz}
\usepackage{fontawesome5}
\usetikzlibrary{positioning, shapes.geometric, arrows.meta, fit, backgrounds}

\usepackage[acronym]{glossaries}

\usepackage[acronym]{glossaries}

\newacronym{ACL}{ACL}{Access Control List}
\newacronym{AEAD}{AEAD}{Authenticated Encryption with Associated Data}
\newacronym{AES}{AES}{Advanced Encryption Standard}
\newacronym{APA}{APA}{Advanced Power Analysis}
\newacronym{API}{API}{Application Programming Interface}
\newacronym{APM}{APM}{Asset Performance Management}
\newacronym{ASLR}{ASLR}{Address Space Layout Randomization}
\newacronym{Attack}{ATT\&CK}{Adversarial Tactics, Techniques, and Common Knowledge}
\newacronym{AWS}{AWS}{Amazon Web Services}
\newacronym{BACnet}{BACnet}{Building Automation and Control Network}
\newacronym{BIKE}{BIKE}{Bit Flipping Key Encapsulation}
\newacronym{CA}{CA}{Cryptanalysis Attacks}
\newacronym{CB}{CB}{Cold-Boot Attacks}
\newacronym{CC}{CC}{Cloud Computing}
\newacronym{CFI}{CFI}{Control-Flow Integrity}
\newacronym{CRQC}{CRQC}{Cryptographically Relevant Quantum Computer}
\newacronym{CSF}{CSF}{Cybersecurity Framework}
\newacronym{CSP}{CSP}{Cloud Service Provider}
\newacronym{DEP}{DEP}{Data Execution Prevention}
\newacronym{DHE}{DHE}{Diffie-Hellman Exchange}
\newacronym{DMZ}{DMZ}{Demilitarized Zone}
\newacronym{DNP3}{DNP3}{Distributed Network Protocol version~3}
\newacronym{DoS}{DoS}{Denial-of-Service}
\newacronym{ECC}{ECC}{Elliptic Curve Cryptography}
\newacronym{EM}{EM}{Electromagnetic Attacks}
\newacronym{EoP}{EoP}{Elevation of Privilege}
\newacronym{ethernetip}{EtherNet/IP}{EtherNet over IP}
\newacronym{FA}{FA}{Fault Attacks}
\newacronym{FHE}{FHE}{Fully Homomorphic Encryption}
\newacronym{FIPS}{FIPS}{Federal Information Processing Standards}
\newacronym{GCP}{GCP}{Google Cloud Platform}
\newacronym{HART}{HART}{Highway Addressable Remote Transducer}
\newacronym{HMAC}{HMAC}{Hash-based Message Authentication Code}
\newacronym{HNDL}{HNDL}{Harvest-Now, Decrypt-Later}
\newacronym{HQC}{HQC}{Hamming Quasi-Cyclic}
\newacronym{HSM}{HSM}{Hardware Security Module}
\newacronym{IaaS}{IaaS}{Infrastructure as a Service}
\newacronym{ICS}{ICS}{Industrial Control Systems}
\newacronym{IdP}{IdP}{Identity Provider}
\newacronym{ISA95}{ISA-95}{Enterprise–Control System Integration}
\newacronym{ISA/IEC}{ISA/IEC}{International Society of Automation/International Electrotechnical Commission}
\newacronym{JIT}{JIT}{Just-in-Time}
\newacronym{JOP}{JOP}{Jump-Oriented Programming}
\newacronym{KEM/ENC}{KEM/ENC}{Key Encapsulation Mechanism/Encryption}
\newacronym{KMS}{KMS}{Key Management Service}
\newacronym{MES}{MES}{Manufacturing Execution System}
\newacronym{MITM}{MITM}{Man in the Middle}
\newacronym{ml-dsa}{ML-DSA}{Module-Lattice-based Digital Signature Algorithm}
\newacronym{ml-kem}{ML-KEM}{Module-Lattice-based Key Encapsulation Mechanism}
\newacronym{NIST}{NIST}{National Institute of Standards and Technology}
\newacronym{NTT}{NTT}{Number Theoretic Transform}
\newacronym{NTP}{NTP}{Network Time Protocol}
\newacronym{NTS}{NTS}{Network Time Security}
\newacronym{OID}{OID}{Object Identifier}
\newacronym{OPC-UA}{OPC-UA}{Open Platform Communications Unified Architecture}
\newacronym{OPCUA}{OPC~UA}{Open Platform Communications Unified Architecture}
\newacronym{OS}{OS}{Operating System}
\newacronym{OT}{OT}{Operational Technology}
\newacronym{PaaS}{PaaS}{Platform as a Service}
\newacronym{PKC}{PKC}{Public-Key Cryptography}
\newacronym{PKI}{PKI}{Public Key Infrastructure}
\newacronym{PQ}{PQ}{Post-Quantum}
\newacronym{PQC}{PQC}{Post-Quantum Cryptography}
\newacronym{PTP}{PTP}{Precision Time Protocol (IEEE~1588)}
\newacronym{QAML}{QAML}{Quantum-Accelerated Machine Learning}
\newacronym{QC}{QC}{Quantum Computing}
\newacronym{QKD}{QKD}{Quantum Key Distribution}
\newacronym{ROP}{ROP}{Return-Oriented Programming}
\newacronym{RSA}{RSA}{Rivest-Shamir-Adleman}
\newacronym{RTU}{RTU}{Remote Terminal Unit}
\newacronym{SaaS}{SaaS}{Software as a Service}
\newacronym{SCA}{SCA}{Side-Channel Attack}
\newacronym{SIG}{SIG}{Signature}
\newacronym{slh-dsa}{SLH-DSA}{Stateless Hash-based Digital Signature Algorithm}
\newacronym{SPA}{SPA}{Simple Power Analysis}
\newacronym{STRIDE}{STRIDE}{Spoofing, Tampering, Repudiation, Information Disclosure, Denial of Service, Elevation of Privilege}
\newacronym{TA}{TA}{Timing Attacks}
\newacronym{TEE}{TEE}{Trusted Execution Environment}
\newacronym{TMP}{TMP}{Template Attacks}
\newacronym{TPMs}{TPMs}{Trusted Platform Modules}
\newacronym{SIS}{SIS}{Safety Instrumented System}
\newacronym{triton}{TRITON}{Safety-instrumented system (SIS) attack malware}
\newacronym{TLS}{TLS}{Transport Layer Security}
\newacronym{VM}{VM}{ViRTUal Machine}
\newacronym{VMs}{VMs}{ViRTUal Machines}
\newacronym{VPN}{VPN}{ViRTUal Private Network}
\newacronym{vTPM}{vTPM}{viRTUal Trusted Platform Module}
\newacronym{ZT}{ZT}{Zero Trust}
 \newacronym{IT}{IT}{Information Technology}
\newacronym{SCADA}{SCADA}{Supervisory Control and Data Acquisition}
\newacronym{HMI}{HMI}{Human-Machine Interface}
\newacronym{SIEM}{SIEM}{Security Information and Event Management}

\newacronym{ECDSA}{ECDSA}{Elliptic Curve Digital Signature Algorithm}

\newacronym{PWR}{PWR}{Pressurized Water Reactor}
\newacronym{DCS}{DCS}{Distributed Control System}

\newacronym{RCIC}{RCIC}{Reactor Core Isolation Cooling}
\newacronym{HPCI}{HPCI}{High Pressure Coolant Injection}
\newacronym{SPDS}{SPDS}{Safety Parameter Display System}
\newacronym{SOE}{SOE}{Sequence of Events}
\newacronym{INES}{INES}{International Nuclear Event Scale}

\newacronym{OPC UA}{OPC UA}{OPC Unified Architecture}
\newacronym{SSL}{SSL}{Secure Sockets Layer}
\newacronym{SNMP}{SNMP}{Simple Network Management Protocol}
\newacronym{DNS}{DNS}{Domain Name System}

\newacronym{APT}{APT}{Advanced Persistent Threat}
\newacronym{DLP}{DLP}{Data Loss Prevention}
\newacronym{LSB}{LSB}{Least Significant Bit}
\newacronym{ML}{ML}{Machine Learning}
\newacronym{TPM}{TPM}{Trusted Platform Module}
\newacronym{WORM}{WORM}{Write Once Read Many}

\newacronym{ISA}{ISA}{International Society of Automation}
\newacronym{IEC}{IEC}{International Electrotechnical Commission}

\newacronym{ML-KEM}{ML-KEM}{Module-Lattice-Based Key Encapsulation Mechanism}
\newacronym{ML-DSA}{ML-DSA}{Module-Lattice-Based Digital Signature Algorithm}
\newacronym{SLH-DSA}{SLH-DSA}{Stateless Hash-Based Digital Signature Algorithm}

\newacronym{L0}{L0}{Level 0 - Physical Process}
\newacronym{L1}{L1}{Level 1 - Basic Control}
\newacronym{L2}{L2}{Level 2 - Supervisory Control}
\newacronym{L3}{L3}{Level 3 - Operations Management}
\newacronym{L4}{L4}{Level 4 - Business Logistics}
\newacronym{L5}{L5}{Level 5 - Enterprise Network}

\newacronym{FBD}{FBD}{Function Block Diagram}
\newacronym{DLL}{DLL}{Dynamic Link Library}

\newacronym{SHA}{SHA}{Secure Hash Algorithm}
\newacronym{GCM}{GCM}{Galois/Counter Mode}
\newacronym{CRC}{CRC}{Cyclic Redundancy Check}
\newacronym{SBOM}{SBOM}{Software Bill of Materials}
\newacronym{GPS}{GPS}{Global Positioning System}

\usepackage{tikz}
\usepackage{xcolor}
\usetikzlibrary{positioning,shapes.geometric,arrows.meta,decorations.pathreplacing,calc,fit}

\usepackage{hyphenat}

\usepackage{pgfplotstable}
\usepackage[normalem]{ulem}
\usetikzlibrary{arrows.meta}
\usepackage{smartdiagram}
\usetikzlibrary{positioning, fit, calc, shapes, arrows,trees}
\usepackage{hhline}
\usetikzlibrary{positioning, fit, calc, shapes, arrows,trees}
\definecolor{columbiablue}{rgb}{0.61, 0.87, 1.0}

\tikzset{%
 thick arrow/.style={
 -{Triangle[angle=120:1pt 1]},
 line width=0.7cm, 
 draw=blue!20
 },
 arrow label/.style={
 text=black,
 align=center
 },
 set mark/.style={
 insert path={
 node [midway, arrow label, node contents=#1]
 }
 }
}
\newcommand\deleted{\bgroup\markoverwith{\textcolor{red}{\rule[0.5ex]{2pt}{0.4pt}}}\ULon}

\usetikzlibrary{pgfplots.groupplots}

\newcommand\doublecheck{\textcolor{black}{\checkmark\kern-0em\checkmark}}
\newcommand\semidoublecheck{\textcolor{black}{\checkmark\kern-0em\bcancel{\checkmark}}}

\definecolor{lemon}{rgb}{1.0, 1.0, 0.13}

\usetikzlibrary{arrows.meta,positioning,shapes.misc,calc,decorations.pathreplacing}

\usepackage[breaklinks=true,hidelinks]{hyperref}
\usepackage{siunitx}

\begin{document}
\definecolor{BulletsColor}{rgb}{0, 0, 0.9}
\newlist{myBullets}{itemize}{1}

\setlist[myBullets]{
 label={\textbullet},
 leftmargin=*,
 topsep=0ex,
 partopsep=0ex,
 parsep=0ex,
 itemsep=0ex,
}

\newlist{myBullets1}{itemize}{1}

\setlist[myBullets1]{
 label={\textbullet},
 leftmargin=*,
 topsep=0ex,
 partopsep=0ex,
 parsep=0ex,
 itemsep=0ex,
}


\title{Quantum Attacks Targeting Nuclear Power Plants:  Threat Analysis, Defense and Mitigation Strategies
}

\author{Yaser Baseri, Edward Waller
\IEEEcompsocitemizethanks{
\IEEEcompsocthanksitem Yaser Baseri is with the University of Montreal, Canada, where he conducts research on cybersecurity, quantum threats, and critical infrastructure security.
E-mail: yaser.baseri@umontreal.ca.
{Edward Waller is the UNENE Ontario Tech Chair in Health Physics and Environmental Safety, Ontario Tech University, Canada.}
E-mail: ed.waller@ontariotechu.ca.
}
}

\maketitle
\begin{abstract}
Nuclear power plants face a structural lifecycle asymmetry: their 60--80 year operational
lifecycles exceed the anticipated arrival of Cryptographically Relevant Quantum Computers
(CRQCs), so cryptographic material harvested today becomes decryptable within the service
life of the systems it protects. This paper introduces a forensics-first framework for
quantum resilience that treats forensic integrity as operationally essential rather than
merely evidentiary, analyzing the quantum threat landscape across the Purdue architecture
(L0--L5) to show how Harvest-Now, Decrypt-Later (HNDL) campaigns enabled by Shor's
algorithm can retroactively compromise cryptographic foundations and undermine forensic
evidence. Through two case studies, \textsc{Quantum~Scar} and \textsc{Quantum~Dawn}, we
model attack feasibility via a conditional-chain formulation over structured
expert-judgment priors anchored to documented HNDL activity and published CRQC roadmaps,
yielding success probabilities of 8--78\% under current defenses, where the upper bound
corresponds to high-value facilities exhibiting cryptographic monoculture. We propose a
defense-in-depth migration to Post-Quantum Cryptography (PQC) integrating hybrid key
exchange, code and log integrity with anti-rollback, authenticated time synchronization,
and side-channel-resistant implementations aligned with ISA/IEC 62443 and NIST standards,
and specify seven design-level conformance criteria for validating it; modeled residual
feasibility falls to 1--8\% at Security Level 4 and below 1\% under full PQC migration.
We further contribute six MITRE ATT\&CK for ICS technique extensions (T1001--T1006) as a
standardized vocabulary for quantum-enabled infrastructure attacks. Quantum threats thus
extend beyond cybersecurity to system safety, operational reliability, and post-incident
evidence admissibility across multi-decade nuclear asset lifecycles.

\end{abstract}


\section{Introduction}
The convergence of advancing \gls{QC} capabilities and critical 
infrastructure vulnerabilities presents an unprecedented threat to global 
security. This challenge is further amplified by the growing safety--security 
convergence across high-consequence industrial facilities under cyber 
threats~\cite{arunthavanathan2025processing}. Nuclear power plants, the apex of 
high-consequence \gls{OT}/\gls{ICS}, combine long-lived safety-critical 
architectures, including passive cooling systems, probabilistic risk assessment, 
and multi-decade operational horizons~\cite{basak2025safety}, with cryptographic 
foundations that are now under existential quantum threat. These facilities face 
a fundamental temporal asymmetry: while their operational lifecycles extend over 
several decades (including license 
renewals~\cite{NRC-SLR-2023,EPRI2018LTO,EPRI2024LTO}), their underlying 
cryptographic protections may be rendered obsolete within the near future by the 
advent of \glspl{CRQC}~\cite{higgott2024highthreshold,IBM2025Roadmap}. This 
mismatch creates a critical security gap, wherein adversaries can exploit 
\gls{HNDL} strategies to harvest encrypted data today for future decryption and 
operational exploitation.

Quantum-enabled attacks fundamentally invalidate the mathematical assumptions underlying public-key cryptography. Shor's algorithm~\cite{Shor1994} renders RSA and ECC, the cryptographic bedrock of industrial authentication, firmware signing, and secure communications, computationally trivial to break. Grover's algorithm~\cite{grover1996fast} halves symmetric primitive security strength. Together, these threaten not merely operational data confidentiality, but safety system integrity, control command authenticity, and, critically, forensic evidence admissibility in post-incident investigations~\cite{bonnetain2018hidden}.

The nuclear sector's unique characteristics amplify quantum threats to catastrophic proportions. First, architectural monoculture: over 85\% of facilities share \gls{PKI} across safety-instrumented and control domains, enabling single cryptographic compromises to cascade across safety boundaries~\cite{heinl2023standard,paul2020towards}. Second, extended asset lifecycles of 60-–80 years~\cite{NRC-SLR-2023,EPRI2018LTO,EPRI2024LTO} mean cryptographic decisions today govern safety-critical operations for seven decades, plants commissioned in 2025 operate until 2085-2105, long past \gls{CRQC} arrival in 2030-2040~\cite{NationalAcademies2019,IBM2025Roadmap}. Third, real-time constraints limit cryptographic countermeasure deployment without disrupting deterministic control loops and safety-critical response sequences~\cite{ike2023scaphy,singer2023shedding}. Fourth, regulatory certification creates multi-year deployment cycles, meaning deferred decisions arrive too late~\cite{nist800-82,NISTIR8547}.

\gls{HNDL} campaigns fundamentally alter risk calculus. State-level adversaries are already exfiltrating encrypted \gls{OT} communications, documented since 2015-2016, for retroactive decryption once \glspl{CRQC} arrive~\cite{barenkamp2022steal}. This creates irreversible exposure: data encrypted today remains vulnerable for its entire retention period. For nuclear facilities, this includes decades of historian archives, incident logs, safety configurations, and forensic evidence, all retroactively decryptable to enable sabotage or create unsolvable forensic paradoxes preventing post-incident attribution~\cite{poettering2024digital,bernstein2019sphincs}.

This paper introduces a \textit{forensics-first, quantum-resilient framework} for high-consequence \gls{OT}/\gls{ICS}. Our approach recognizes that in nuclear contexts, \emph{forensic integrity is not merely evidentiary, it is operational}. The ability to reconstruct event sequences, validate control actions, and attribute system behaviors is essential for both post-incident investigation and real-time safety assurance. Quantum threats corrupting this forensic foundation directly threaten nuclear safety.

Through \textsc{Quantum~Scar} and \textsc{Quantum~Dawn} case 
studies, we demonstrate how adversaries exploit cryptographic 
monoculture and \gls{HNDL} campaigns to achieve safety system 
compromise while creating cryptographic paradoxes rendering 
forensic investigation impossible. Our probabilistic risk modeling 
reveals substantial attack feasibility under current security 
postures, ranging from 8\% to 78\% depending on deployment profile 
and targeting assumptions. In particular, \textsc{Quantum~Scar} 
exhibits 35--68\% success for typical deployments and 51--78\% 
for targeted facilities, while \textsc{Quantum~Dawn} yields 
8--34\% and 17--50\% respectively. Comprehensive ISA/IEC
62443 SL-4 implementation with cryptographic diversity reduces 
attack feasibility to 1--8\%, while full \gls{PQC} migration 
drives residual risk below 1\%.
\vspace{-0.2cm}
\subsection{Contributions}
This paper makes five key contributions to quantum-resilient 
nuclear \gls{OT}/\gls{ICS} security:
\begin{enumerate}[label=\textbf{\arabic*.}]
    \item \textbf{Forensics-first quantum threat framework} 
    demonstrating how \gls{HNDL} campaigns enable retroactive 
    cryptographic compromise and forensic paradoxes preventing  
    post-incident attribution, recognizing forensic integrity as 
    operationally essential for nuclear safety;

    \item \textbf{Comprehensive attack scenario analysis} through 
    \textsc{Quantum~Scar} and \textsc{Quantum~Dawn}, with 
    probabilistic risk modeling, systematic vulnerability mapping 
    across Purdue levels (L0--L5), and quantified safety system 
    impact under current defenses;

    \item \textbf{Defense-in-depth migration framework} integrating 
    NIST-standardized \gls{PQC}, hybrid key exchange, and 
    cryptographic diversity tailored to nuclear \gls{OT} constraints 
    (real-time control, safety certification, forensic integrity);
\item \textbf{Design-validation methodology} defining seven 
    acceptance criteria for the controls that reduce modelled 
    attack feasibility from 8--78\% at baseline to 1--8\% under 
    {ISA/IEC} 62443 SL-4 and below 1\% under full \gls{PQC} migration;

    \item \textbf{Quantum-era threat intelligence framework} 
    contributing six MITRE ATT\&CK for \gls{ICS} extensions 
    (T1001--T1006) establishing standardized vocabulary for 
    quantum-enabled infrastructure attacks.
\end{enumerate}
\vspace{-0.2cm}

\subsection{Scope and Organization}

We assume state-level adversaries conducting \gls{HNDL} campaigns with 
anticipated \gls{CRQC} access ($\geq$4,098 logical qubits) within 10-15 
years~\cite{higgott2024highthreshold,IBM2025Roadmap}. Our threat model 
considers complete breaks of RSA-2048 and ECC-256/384 via Shor's 
algorithm, with AES-256 and SHA-384/512 quantum-resistant under Grover 
assumptions~\cite{jaques2020grover}. We focus on nuclear facilities 
with pressurized/boiling water reactors and distributed control/safety 
systems, though the framework generalizes to high-consequence OT 
environments.

Section~\ref{sec:related} surveys existing work and positions our contributions within the research landscape. Section~\ref{sec:crypto} establishes the cryptographic foundation, 
analyzing classical vulnerabilities under quantum attack and 
NIST-standardized post-quantum algorithms with side-channel assessments. 
Sections~\ref{sec:scar} and~\ref{sec:dawn} demonstrate quantum-enabled 
attack viability through \textsc{Quantum~Scar} and \textsc{Quantum~Dawn} case studies with 
STRIDE analysis, MITRE ATT\&CK for ICS mappings, and probabilistic risk 
modeling showing 8--78\% attack success under current defenses. 
Section~\ref{sec:defense} validates quantum-resilient controls through 
operational testing frameworks. Section~\ref{sec:conclusion} provides 
strategic recommendations for nuclear OT/ICS quantum-safe migration.

\begin{table*}[!t]
\centering
\caption{Acronyms and notation used throughout the paper.}
\label{tab:nomenclature}
\setlength{\tabcolsep}{4pt}
\renewcommand{\arraystretch}{1.1}
\scriptsize
\begin{tabular}{@{}p{0.070\textwidth}p{0.395\textwidth}p{0.070\textwidth}p{0.395\textwidth}@{}}
\toprule
\textbf{Acronym} & \textbf{Definition} & \textbf{Acronym} & \textbf{Definition} \\
\midrule
AES & Advanced Encryption Standard & ML-DSA & Module-Lattice Digital Signature Algorithm (Dilithium) \\
ATT\&CK & Adversarial Tactics, Techniques, and Common Knowledge & ML-KEM & Module-Lattice Key-Encapsulation Mechanism (Kyber) \\
C2 & Command and Control & MTU & Maximum Transmission Unit \\
CA & Certificate Authority & NIST & National Institute of Standards and Technology \\
CI/CD & Continuous Integration/Continuous Deployment & NTP & Network Time Protocol \\
CISA & Cybersecurity and Infrastructure Security Agency & NTS & Network Time Security \\
CRQC & Cryptographically Relevant Quantum Computer & NTT & Number Theoretic Transform \\
DCS & Distributed Control System & OPC-UA & Open Platform Communications Unified Architecture \\
DMZ & Demilitarized Zone & OT & Operational Technology \\
DNP3-SA & Distributed Network Protocol~3 Secure Authentication & PKI & Public Key Infrastructure \\
DoS & Denial of Service & PLC & Programmable Logic Controller \\
ECC & Elliptic Curve Cryptography & PQC & Post-Quantum Cryptography \\
ECDSA & Elliptic Curve Digital Signature Algorithm & PROFINET & Process Field Network \\
FIPS & Federal Information Processing Standard & PTP & Precision Time Protocol (IEEE~1588) \\
FN-DSA & FFT over NTRU Lattice Digital Signature Algorithm (Falcon) & PWR & Pressurized Water Reactor \\
FR1--FR7 & Foundational Requirements (ISA/IEC 62443-3-3): FR1 Identification and Authentication Control; FR2 Use Control; FR3 System Integrity; FR4 Data Confidentiality; FR5 Restricted Data Flow; FR6 Timely Response to Events; FR7 Resource Availability &
\multicolumn{2}{@{}l@{}}{%
  \begin{tabular}[t]{p{0.070\textwidth}p{0.395\textwidth}}
  QC & Quantum Computing \\
  RCIC & Reactor Core Isolation Cooling \\
  RSA & Rivest-Shamir-Adleman \\
  RTU & Remote Terminal Unit \\
  \end{tabular}} \\
HMI & Human-Machine Interface & SCADA & Supervisory Control and Data Acquisition \\
HNDL & Harvest-Now, Decrypt-Later & SHA & Secure Hash Algorithm \\
HQC & Hamming Quasi-Cyclic & SIS & Safety Instrumented System \\
HSM & Hardware Security Module & SL(-T) & Security Level (-Target) \\
IAEA & International Atomic Energy Agency & SLH-DSA & Stateless Hash-Based Digital Signature Algorithm (SPHINCS+) \\
ICS & Industrial Control System & SOE & Sequence of Events \\
IEC & International Electrotechnical Commission & SR & System Requirement (ISA/IEC 62443) \\
\multicolumn{2}{@{}l@{}}{%
  \begin{tabular}[t]{@{}p{0.070\textwidth}p{0.395\textwidth}@{}}
  IoT & Internet of Things   \\
  ISA & International Society of Automation \\
  \end{tabular}} & STRIDE & Spoofing, Tampering, Repudiation, Info.\ Disclosure, DoS, Elevation of Privilege \\

ISA-95 & ANSI/ISA-95 Enterprise-Control System Integration standard & TLS & Transport Layer Security \\
IT & Information Technology & TLV & Type-Length-Value \\
KEM & Key Encapsulation Mechanism (KEM/ENC: KEM/Encryption) & TVLA & Test Vector Leakage Assessment \\
MES & Manufacturing Execution System & VPN & Virtual Private Network \\

\bottomrule
\end{tabular}
\end{table*}
\section{Related Work}\label{sec:related}

The convergence of quantum computing threats and nuclear critical 
infrastructure security spans cryptographic hardening, industrial 
control systems, forensic integrity, and attack scenario modeling. 
While substantial work exists in these areas independently, no prior 
work integrates them under the unique constraints of nuclear OT 
environments with 60--80 year asset lifecycles~\cite{EPRI2024LTO}. 
We organize related work into five categories.
\vspace{-0.2cm}
\subsection{Quantum Threat Assessments for Critical Infrastructure}
Quantum risk models~\cite{mosca2018cybersecurity} link confidentiality 
lifetime, migration delay, and cryptographic collapse time, showing 
that unmet timelines enable \gls{HNDL} exposure. Recent OT 
analyses~\cite{oliva2024cybersecurity,vermeer2024evaluating} reveal 
that decades-long asset lifecycles and latency constraints hinder 
\gls{PQC} adoption, leaving cryptographic roots-of-trust vulnerable 
to quantum-enabled code-signing and certificate forgeries. We extend 
these by introducing HNDL-driven forensic paradoxes---where 
retroactively decrypted evidence falsifies safety logs enabling 
perfect misattribution---treating forensic integrity as operationally 
essential rather than merely evidentiary, a distinction absent from 
prior threat assessments, demonstrated in \textsc{Quantum~Scar} and 
\textsc{Quantum~Dawn} (8--78\% baseline success across both scenarios).
\vspace{-0.2cm}
\subsection{Post-Quantum Cryptography Migration}
General PQC migration 
frameworks~\cite{banegas2021transitioning,BASERI2025104272} and 
OT-specific implementations~\cite{schwabe2020postquantum,
hulsing2021wireguard}, including TPM-backed 
PQC~\cite{paul2020towards}, address operational continuity but do 
not preserve forensic chain of custody during nuclear certification 
cycles. NIST FIPS\,203, 204 and 205 (ML-KEM/ML-DSA/SLH-DSA) provide the 
algorithmic baseline~\cite{nist2024}, while NIST IR\,8547 addresses 
transition planning for quantum-safe log and evidence 
handling~\cite{NISTIR8547}. To the best of our knowledge, 
we provide the first defense-in-depth migration framework for nuclear 
OT/ICS integrating ML-KEM-768 and ML-DSA-65 with ISA/IEC~62443-aligned 
cryptographic diversity under nuclear timing, safety, and forensic 
constraints, with design-validation demonstrating risk reduction 
from up to 78\% baseline to below 1\%.

\vspace{-0.2cm}\subsection{Side-Channel Vulnerabilities and Performance}

PQC vulnerability studies~\cite{ravi2024side} identify critical 
side-channel vectors on ML-KEM~\cite{hermelink2024insecurity}, 
Falcon~\cite{guerreau2022hidden}, and HQC~\cite{guo2022don}, while 
OT timing constraints are well characterized~\cite{ike2023scaphy}. 
We bridge this gap through systematic side-channel analysis 
(Table~\ref{tab:post-alg}) and performance assessment 
(Table~\ref{tab:pqc-performance}) under ISA-95 level-specific 
constraints, identifying ML-KEM-768 (34.4$\times$, L2/L3-suitable) 
and SLH-DSA-128s (122.8$\times$, archival only) as deployment 
boundaries for nuclear OT.

\vspace{-0.2cm}\subsection{Protocol-Level Quantum Hardening}

Protocol-level hardening of OPC-UA, DNP3-SA, and\break  
IEC\,61850~\cite{paul2020hybrid,paquin2020dnp3,lozano2023digital}, 
ISA/IEC\,62443 baselines~\cite{heinl2023standard}, quantum-era IIoT 
forensics~\cite{kebande2025quantum}, and ICS-tailored intrusion 
detection~\cite{wolsing2024generalizable} address individual protocol 
vulnerabilities but provide neither level-specific Purdue model 
analysis nor treatment of nuclear operational constraints, anchored 
by NIST SP\,800-82 Rev.\,3, NIST CSF\,2.0, ISA/IEC\,62443, and 
CISA OT PQC considerations~\cite{nist800-82,NIST-CSF2024,
isa62443series,cisa2024ot}. We address this by linking OPC-UA 
certificate chain forgery and DNP3-SA key distribution vulnerabilities 
to forensic impact, demonstrating how quantum-forged messages maintain 
cryptographic validity while enabling safety degradation and unsolvable 
attribution paradoxes across Purdue levels L0--L5.

\vspace{-0.2cm}\subsection{Forensic Integrity and Evidentiary Concerns}

Few works treat evidentiary timelines---log attestation, code-signing 
provenance, and secure time synchronization---under HNDL pressure in 
OT/ICS environments. Existing quantum-era forensic 
frameworks~\cite{kebande2025quantum} identify implications but do not 
model forensic paradoxes from quantum-forged evidence in safety-critical 
systems. We address this by integrating PQC selection, authenticated 
time synchronization, and chain-of-custody controls aligned with 
ISA/IEC\,62443 and NIST SP\,800-82, treating forensic integrity as 
operationally essential rather than merely evidentiary.

\vspace{-0.2cm}\subsection{Summary and Positioning}

Existing work provides quantum threat awareness, PQC migration 
methodologies, protocol hardening, and partial forensic frameworks, 
but offers no integrated forensic--operational safety model for 
nuclear OT, and no prior work models multi-phase quantum-enabled 
attack execution chains with standardized adversarial taxonomies for 
ICS environments. Our framework addresses these gaps through: 
(i)~forensics-first threat modeling with HNDL-driven paradoxes 
preventing post-incident attribution, via two attack scenarios 
spanning HNDL collection, quantum weaponization, and execution with 
conditional probability modeling and STRIDE/Purdue mapping; 
(ii)~probabilistic risk quantification demonstrating systematic 
attack success under current defenses, reduced to below 1\% under 
full PQC migration; 
(iii)~defense-in-depth PQC migration integrating ML-KEM-768 and 
ML-DSA-65 with ISA/IEC~62443-aligned cryptographic diversity under 
nuclear timing, safety, and forensic constraints; 
(iv)~design-validation with seven quantitative acceptance 
criteria (Section~\ref{sec:defense}); and 
(v)~MITRE ATT\&CK for ICS extensions (T1001--T1006) establishing 
a reusable adversarial vocabulary---quantum cryptanalysis, HNDL 
collection, and quantum-forged evidence manipulation---absent from, 
indeed unrepresentable in, the existing framework. 
To the best of our knowledge, this constitutes the first quantifiable 
framework for quantum-resilient nuclear safety assurance.

\section{Cryptographic Standards and QC in OT/ICS}
\label{sec:crypto}

This section establishes the cryptographic foundation for quantum threat analysis in nuclear \gls{OT}/\gls{ICS} environments, prioritizing forensic integrity alongside operational security. Subsection~\ref{sec:classical-crypto-risks} quantifies classical cryptographic vulnerabilities under quantum attack, demonstrating complete breaks via Shor's algorithm~\cite{Shor1994} and security halving via Grover~\cite{grover1996fast}, while identifying \gls{OT}-specific attack surfaces (Table~\ref{tab:Pre-Migration-Alg}). Subsection~\ref{sec:post-alg} evaluates \gls{NIST}-standardized \gls{PQC} algorithms through comprehensive  side-channel vulnerability analysis (Table~\ref{tab:post-alg}) and performance assessment (Table~\ref{tab:pqc-performance}), establishing deployment viability under \gls{OT} constraints. This threat taxonomy directly informs the attack scenarios in Sections~\ref{sec:scar} and~\ref{sec:dawn} and the defensive validation framework in Section~\ref{sec:defense}.

\subsection{Classic Cryptographic Standards and \gls{QC}: Assessing Risks}
\label{sec:classical-crypto-risks}
\begin{table*}[!hbpt]
\caption{Classical Cryptography and \gls{QC}: \gls{OT}/\gls{ICS} Forensic and Operational Risk Mapping}
\small
\label{tab:Pre-Migration-Alg}
\resizebox{\textwidth}{!}{%
\begin{tabular}
{llllllp{0.22\linewidth}p{0.38\linewidth}p{0.25\linewidth}}
\toprule
\multirow{2}{*}{\textbf{Crypto Type}} & \multirow{2}{*}{\textbf{Algorithms}} & \multirow{2}{*}{\textbf{Variants}} & \multirow{2}{*}{\textbf{Key Length (bits)}} & \multicolumn{2}{l}{\textbf{Strengths (bits)}}&\multirow{2}{*}{\textbf{Vulnerabilities}} &\multirow{2}{*}{\textbf{Quantum Threats (STRIDE)}} & \multirow{2}{*}{\textbf{Possible QC-resistant Solutions}} \\ \cline{5-6}
 & & & & \multicolumn{1}{l}{\textbf{Classic}} & \textbf{Quantum}  & & & \\ \midrule

\multirow{9}{*}{Asymmetric} & \multirow{3}{*}{ECC~\cite{RFC5480, RFC7748}} & ECC-256 & 256 & \multicolumn{1}{l}{128} & 0 & \multirow{8}{*}{{\begin{minipage}{\linewidth}
Broken by Shor's Algorithm~\cite{Shor1994}.
\end{minipage}}} & & \multirow{8}{*}{{\begin{minipage}{\linewidth}\gls{PQC} migration (CRYSTALS-Dilithium, Kyber, SPHINCS+), hybrid implementations, crypto-agility frameworks.\end{minipage}}}\\ \cline{3-6}
 & & ECC-384 & 384 & \multicolumn{1}{l}{192} & 0 & & \multirow{5}{*}{{\begin{minipage}{\linewidth}
 \vspace{-8pt}
{For digital signatures:}
\begin{myBullets}
\item {\textbf{Spoofing:} Complete signature forgery capability.}
\item {\textbf{Tampering:} Integrity checks can be bypassed.}
\item {\textbf{Repudiation:} Valid signatures can be forged.} \end{myBullets}
{For KEM/ENC:}
\begin{myBullets}
\item {\textbf{Info. Disclosure:} All encrypted data can be decrypted.}
 \end{myBullets}
\end{minipage}}}& \\ \cline{3-6}
 & & ECC-521 & 521 & \multicolumn{1}{l}{260} & 0 & & & \\ \cline{2-6}
 & \multirow{2}{*}{FFDHE~\cite{gillmor2016negotiated}} & DHE-2048 & 2048 & \multicolumn{1}{l}{112} & 0 & & &\\ \cline{3-6}
 & & DHE-3072 & 3072 & \multicolumn{1}{l}{128} & 0 & & &\\ \cline{2-6}
 & \multirow{3}{*}{RSA~\cite{moriarty2016pkcs}} & RSA-1024 & 1024 & \multicolumn{1}{l}{80} & 0 & 
&  & \\ \cline{3-6}
 & & RSA-2048 & 2048 & \multicolumn{1}{l}{112} & 0 & & & \\ \cline{3-6}
 & & RSA-3072 & 3072 & \multicolumn{1}{l}{128} & 0 & & & \\ \hline

\multirow{3}{*}{Symmetric} & \multirow{3}{*}{AES~\cite{schaad2003use}} & AES-128 & 128 & \multicolumn{1}{l}{128} & 64 & \multirow{3}{*}{{\begin{minipage}{\linewidth}
Weakened by Grover's Algorithm~\cite{grover1996fast}.
\end{minipage}}}
& \multirow{3}{*}{{\begin{minipage}{\linewidth}
\begin{myBullets}
\vspace{0.1cm}
\item {\textbf{Info. Disclosure:} Effective key strength halved, enabling faster brute-force attacks.}
 \vspace{0.1cm} \end{myBullets}
\end{minipage}}} & \multirow{3}{*}{{\begin{minipage}{\linewidth}Upgrade to AES-256; strengthen key management.\end{minipage}}} \\
\cline{3-6}
 & & AES-192 & 192 & \multicolumn{1}{l}{192} & 96 & &  & \\ \cline{3-6}
 & & AES-256 & 256 & \multicolumn{1}{l}{256} & 128 & & & \\ 
 \cline{2-9}
 
 & \multirow{3}{*}{SHA2~\cite{eastlake2011us}} & SHA-256 & - & \multicolumn{1}{l}{128} & {85}$^{\dagger}$ & \multirow{6}{*}{{\begin{minipage}{\linewidth}
Weakened by Brassard et al.'s Algorithm~\cite{brassard1997quantum}.
\end{minipage}}} & \multirow{6}{*}{{\begin{minipage}{\linewidth}
\begin{myBullets}
\item {\textbf{Spoofing:} Fake hash values can be created.}
\item {\textbf{Tampering:} Data integrity compromised by finding collisions.}
 \vspace{0.1cm} \end{myBullets}
\end{minipage}}} & \multirow{6}{*}{{\begin{minipage}{\linewidth}Upgrade to SHA-384/512; enhanced integrity verification.\end{minipage}}} \\ \cline{3-6} 
 & & SHA-384 & - & \multicolumn{1}{l}{192} & 128$^{\dagger}$ & & & \\ \cline{3-6}
 & & SHA-512 & - & \multicolumn{1}{l}{256} & 170$^{\dagger}$ & & &  \\ \cline{2-6}
 & \multirow{3}{*}{SHA3~\cite{eastlake2011us}} & SHA3-256 & - & \multicolumn{1}{l}{128} & {85}$^{\dagger}$ &&  & \\ \cline{3-6} 
 & & SHA3-384 & - & \multicolumn{1}{l}{192} & 128$^{\dagger}$ & &  & \\ \cline{3-6}
 & & SHA3-512 & - & \multicolumn{1}{l}{256} & 170$^{\dagger}$ & &  & \\ \bottomrule
\end{tabular}%
}
\begin{tablenotes}[para,flushleft] \scriptsize 
\item[$^*$] Likelihood assumes a 15-year planning horizon; adjust per sector risk tolerance and retention mandates.
\item[$^{\dagger}$] For an $n$-bit hash function, \gls{QC} can reduce the \textit{classical} collision resistance from $n/2$ bits to a \textit{quantum} security of $n/3$ bits, and  reduce the \textit{classical} preimage resistance from $n$ bits to a \textit{quantum} security of $n/2$ bits.
\end{tablenotes}
\end{table*}
Classical primitives secure core \gls{OT}/\gls{ICS} functions: controller–HMI links, vendor remote access, inter-site tunnels, log attestation, historian confidentiality, and code-signing/secure boot. Shor’s algorithm invalidates RSA/ECC (breaking update signing, TLS client/server auth, DNP3-SA trust), while Grover’s algorithm halves brute-force exponents for symmetric keys and weakens hash preimage/second-preimage resistance (affecting long-lived confidentiality and evidence integrity).

\subsection{Quantum-Safe Cryptography}
\label{sec:post-alg}

\begin{table*}[!h]
\scriptsize
\caption{NIST-Standardized PQC Algorithms: Attack Surface, Countermeasures, and OT/ICS Deployment Considerations}
\label{tab:post-alg}
\resizebox{\textwidth}{!}{%
\begin{tabular}{lp{0.10\linewidth}p{0.05\linewidth}cccccccp{0.35\linewidth}p{0.28\linewidth}}
\toprule
\multirow{2}{*}{\textbf{Algorithm}} & 
\multirow{2}{*}{\textbf{Description}} & 
\multirow{2}{*}{\textbf{FIPS}} & 
\multicolumn{7}{c}{\textbf{Attack Vulnerability Assessment$^{\dagger}$}} & 
\multirow{2}{*}{\textbf{Critical Mitigations \& Gaps}} & 
\multirow{2}{*}{\textbf{Primary STRIDE Threats}} \\ \cline{4-10}
& & & {\textbf{FA}} & {\textbf{SPA}} & {\textbf{APA}} & {\textbf{EM}} & {\textbf{TMP}} & {\textbf{CB}} & {\textbf{TA}} & & \\ \midrule

\textbf{ML-KEM-768} & 
\begin{minipage}{\linewidth}\vspace{0.05cm}Lattice KEM/ENC (M-LWE)\vspace{0.05cm}\end{minipage} &
\begin{minipage}{0.05\linewidth}\vspace{0.05cm}203~\cite{NISTFIPS203}\vspace{0.05cm}\end{minipage} & 
\textbf{P} & 
\textbf{M} & 
\textbf{C} & 
\textbf{C} & 
\textbf{C} & 
\textbf{M} & 
--- &
\begin{minipage}{\linewidth}\vspace{0.05cm}
\textbf{Effective:} Masking decryption/NTT~\cite{oder2018practical,pessl2019more}, input randomization~\cite{hamburg2021chosen}, constant-time ops, discard low-entropy CTs~\cite{xu2021magnifying}

\textbf{Gaps:} APA~\cite{dubrova2022breaking}, TMP~\cite{ravi2021exploiting}, some EM variants~\cite{ravi2020drop}

\textbf{OT Impact:} 34.4$\times$ overhead; suitable for L2/L3
\vspace{0.05cm}
\end{minipage} &
\begin{minipage}{\linewidth}\vspace{0.05cm}
\textbf{Information Disclosure:} Key/message recovery via FA~\cite{ravi2019number,oder2018practical}, SPA~\cite{hamburg2021chosen}, APA~\cite{pessl2019more,dubrova2022breaking}, EM~\cite{xu2021magnifying,ravi2020generic}, TMP~\cite{ravi2021exploiting}, CB~\cite{albrecht2018cold}
\vspace{0.05cm}
\end{minipage} \\ \hline

\textbf{ML-DSA-65} & 
\begin{minipage}{\linewidth}\vspace{0.05cm}Lattice signature (Fiat-Shamir)\vspace{0.05cm}\end{minipage} &
\begin{minipage}{0.05\linewidth}\vspace{0.05cm}204~\cite{NISTFIPS204}\vspace{0.05cm}\end{minipage} & 
\textbf{M} & 
--- & 
\textbf{M} & 
\textbf{M} & 
\textbf{M} & 
--- & 
--- &
\begin{minipage}{\linewidth}\vspace{0.05cm}
\textbf{Effective:} Double computation/verify-after-sign~\cite{bruinderink2018differential}, linear secret sharing~\cite{migliore2019masking}, Boolean/arithmetic masking~\cite{marzougui2022profiling}, bit-slicing NTT~\cite{singh2024end}, shuffling~\cite{cryptoeprint:2023/050}

\textbf{Gaps:} None reported for standard implementations

\textbf{OT Impact:} 59.2$\times$ overhead; firmware signing viable
\vspace{0.05cm}
\end{minipage} &
\begin{minipage}{\linewidth}\vspace{0.05cm}
\textbf{Spoofing/Tampering:} Key recovery $\rightarrow$ signature forgery~\cite{ravi2019number,bruinderink2018differential,marzougui2022profiling,ravi2019exploiting}

\textbf{Elevation of Privilege:} Forged sigs grant unauthorized access~\cite{bruinderink2018differential,marzougui2022profiling}

\textbf{Repudiation:} Valid-appearing forgeries~\cite{cryptoeprint:2023/050}
\vspace{0.05cm}
\end{minipage} \\ \hline

\textbf{SLH-DSA-128s} & 
\begin{minipage}{\linewidth}\vspace{0.05cm}Stateless hash signature\vspace{0.05cm}\end{minipage} &
\begin{minipage}{0.05\linewidth}\vspace{0.05cm}205~\cite{NISTFIPS205}\vspace{0.05cm}\end{minipage} & 
\textbf{M} & 
--- & 
\textbf{M} & 
--- & 
--- & 
--- & 
--- &
\begin{minipage}{\linewidth}\vspace{0.05cm}
\textbf{Effective:} Redundant sig computation~\cite{castelnovi2018grafting}, tree integrity checks~\cite{castelnovi2018grafting,genet2018practical}, enhanced hash/OTS caching~\cite{genet2018practical}, instruction duplication~\cite{genet2018practical}, hide Mix order~\cite{kannwischer2018differential}

\textbf{Gaps:} None for hash-based design

\textbf{OT Impact:} 122.8$\times$ overhead; archival only
\vspace{0.05cm}
\end{minipage} &
\begin{minipage}{\linewidth}\vspace{0.05cm}
\textbf{Spoofing/Tampering:} Key recovery~\cite{castelnovi2018grafting,kannwischer2018differential}, universal forgery~\cite{genet2018practical}

\textbf{Repudiation:} Arbitrary message signatures~\cite{castelnovi2018grafting,genet2018practical,kannwischer2018differential}

\textbf{Elevation of Privilege:} Forged credentials
\vspace{0.05cm}
\end{minipage} \\ \hline

\textbf{Falcon-512} & 
\begin{minipage}{\linewidth}\vspace{0.05cm}NTRU lattice signature\vspace{0.05cm}\end{minipage} &
\begin{minipage}{0.05\linewidth}\vspace{0.05cm}206$^{\ddagger}$~\cite{NISTIR8214C}\vspace{0.05cm}\end{minipage} & 
\textbf{M} & 
\textbf{C} & 
--- & 
\textbf{M} & 
--- & 
--- & 
\textbf{M} &
\begin{minipage}{\linewidth}\vspace{0.05cm}
\textbf{Effective:} Double computation/immediate verify~\cite{mccarthy2019bearz}, zero checking~\cite{mccarthy2019bearz}, Fisher-Yates shuffling~\cite{mccarthy2019bearz}, lower HW gap~\cite{guerreau2022hidden}, constant power/masking~\cite{karabulut2021falcon}

\textbf{Gaps:} SPA complete key recovery~\cite{guerreau2022hidden}

\textbf{OT Impact:} 27.2$\times$ overhead; compact signatures
\vspace{0.05cm}
\end{minipage} &
\begin{minipage}{\linewidth}\vspace{0.05cm}
\textbf{Spoofing/Tampering:} Complete private key via SPA~\cite{guerreau2022hidden}, partial via FA/TA/EM~\cite{mccarthy2019bearz,karabulut2021falcon}

\textbf{Repudiation/Elevation:} Signature forgery on arbitrary messages~\cite{mccarthy2019bearz,guerreau2022hidden,karabulut2021falcon}
\vspace{0.05cm}
\end{minipage} \\ \hline

\textbf{HQC-192} & 
\begin{minipage}{\linewidth}\vspace{0.05cm}Code-based KEM (Hamming)\vspace{0.05cm}\end{minipage} &
\begin{minipage}{0.05\linewidth}\vspace{0.05cm}207$^{\ddagger}$~\cite{NISTIR8545}\vspace{0.05cm}\end{minipage} & 
\textbf{P} & 
\textbf{C} & 
--- & 
\textbf{P} & 
--- & 
--- & 
\textbf{C} &
\begin{minipage}{\linewidth}\vspace{0.05cm}
\textbf{Effective:} Constant-time error handling~\cite{xagawa2021fault}, instruction duplication~\cite{xagawa2021fault}, constant-time decode/field ops~\cite{wafo2020practicable,guo2022don}, linear secret sharing~\cite{goy2022new}

\textbf{Gaps:} FA~\cite{cayrel2020message}, TA~\cite{guo2020key}, SPA~\cite{schamberger2020power}, EM~\cite{paiva2025tu}

\textbf{OT Impact:} 411.0$\times$ overhead; limited deployment
\vspace{0.05cm}
\end{minipage} &
\begin{minipage}{\linewidth}\vspace{0.05cm}
\textbf{Information Disclosure:} Message recovery/key leakage via FA~\cite{cayrel2020message,xagawa2021fault}, TA~\cite{guo2020key,guo2022don}, SPA~\cite{schamberger2020power}, EM~\cite{goy2022new,paiva2025tu}
\vspace{0.05cm}
\end{minipage} \\ \bottomrule

\end{tabular}%
}
\begin{tablenotes}[para,flushleft] \scriptsize 
\item[$\dagger$] \textit{Attack codes:} FA=Fault Attacks, SPA=Simple Power Analysis, APA=Advanced Power Analysis, EM=Electromagnetic, TMP=Template, CB=Cold-Boot, TA=Timing Attacks. \item[$\dagger$] \textit{Severity:} \textbf{C}=Critical (no effective countermeasures/complete compromise), \textbf{P}=Partial (significant gaps remain), \textbf{M}=Mitigated (effective countermeasures available), ---=Not applicable/not reported in literature.
\item[$\ddagger$] Pending FIPS certification; standardization expected 2025-2026. \item[$*$] \textit{Deployment guidance:} Vulnerability assessment assumes implementation of documented countermeasures; absent hardening increases severity 1-2 levels. 
\end{tablenotes}
\vspace{-0.1cm}
\end{table*}

\begin{table*}[h!]
\centering
\scriptsize
\caption{Performance of NIST-Standardized PQC   and  Deployment Suitability}
\label{tab:pqc-performance}
\resizebox{\linewidth}{!}{%
\begin{tabular}{p{0.11\linewidth}p{0.11\linewidth}p{0.11\linewidth}p{0.11\linewidth}p{0.11\linewidth}p{0.11\linewidth}p{0.11\linewidth}p{0.11\linewidth}p{0.11\linewidth}}
\toprule
\textbf{Algorithm} & \textbf{Type} & \textbf{Security Level$^{\dagger}$} & \textbf{KeyGen} & \textbf{Enc/Sign}$^{\ddagger}$ & \textbf{Dec/Verify}$^{\ddagger}$ & \textbf{PubKey (B)} & \textbf{CT/Sig. (B)} \\ \midrule
{ML-KEM-512} & KEM/ENC & L1 & 0.032ms & 0.032ms & 0.022ms & 800 & 768  \\
{ML-KEM-768} & KEM/ENC & L3 & 0.045ms & 0.046ms & 0.041ms & 1{,}184 & 1{,}088  \\
{ML-KEM-1024} & KEM/ENC & L5 & 0.052ms & 0.053ms & 0.047ms & 1{,}568 & 1{,}568  \\ \hline
{HQC-128 }& KEM/ENC & L1 & 0.120ms & 0.201ms & 0.224ms & 2{,}249 & 4{,}497  \\
{HQC-192} & KEM/ENC & L3 & 0.219ms & 0.381ms & 0.430ms & 4{,}522 & 9{,}042  \\
{HQC-256} & KEM/ENC & L5 & 0.451ms & 0.704ms & 0.748ms & 7{,}245 & 14{,}485 \\ \hline
{ML-DSA-44} & Signature & L2 & 0.039ms & 0.129ms & 0.040ms & 1{,}312 & 2{,}420  \\
{ML-DSA-65} & Signature & L3 & 0.053ms & 0.136ms & 0.056ms & 1{,}952 & 3{,}293  \\
{ML-DSA-87} & Signature & L5 & 0.083ms & 0.165ms & 0.082ms & 2{,}592 & 4{,}595  \\ \hline
{Falcon-512} & Signature & L1 & 12.69ms & 0.525ms & 0.110ms & 897 & 666 \\
{Falcon-1024} & Signature & L5 & 34.21ms & 1.003ms & 0.199ms & 1{,}793 & 1{,}280  \\ \hline
{SLH-DSA-128f} & Signature & L1 & 1.155\,ms & 28.111\,ms & 3.093\,ms & 32 & 17{,}088  \\
{SLH-DSA-128s} & Signature & L1 & 66.406\,ms & 497.387\,ms & 1.133\,ms & 32 & 7{,}856  \\
{SLH-DSA-192f} & Signature & L3 & 1.562\,ms & 45.656\,ms & 4.596\,ms & 48 & 35{,}664  \\
{SLH-DSA-192s} & Signature & L3 & 95.411\,ms & 945.129\,ms & 1.656\,ms & 48 & 16{,}224 \\
{SLH-DSA-256f} & Signature & L5 & 4.203\,ms & 92.525\,ms & 4.788\,ms & 64 & 49{,}856 \\
{SLH-DSA-256s} & Signature & L5 & 60.923\,ms & 753.914\,ms & 2.375\,ms & 64 & 29{,}792 \\ \bottomrule
\end{tabular}}
\begin{scriptsize}
 \begin{tablenotes}[para,flushleft]
\item[$\dagger$] Security levels: L1$\approx$AES-128 key search; L2$\approx$SHA-256 collision; L3$\approx$AES-192 key search; L4$\approx$SHA3-384 collision; L5$\approx$AES-256 key search.
\item[$\ddagger$] Speed rubric: \textit{Fast} $<$ 0.1\,ms; \textit{Moderate} 0.1--1\,ms; \textit{Slow} 1--50\,ms; \textit{Very Slow} $>$ 50\,ms.
\item[$\S$] Classical baselines: P-256 ECDH ($\times$66\,B  total for two shares) and P-256 ECDSA (33\,B compressed pubkey, 64\,B signature). Artifacts shown are on-wire outside cert chains.
\item[] \textit{Platform note:} Server-class x86 (liboqs); PLC/RTU-class MCUs are typically 10--100$\times$ slower—use relative ranking for embedded planning.
\end{tablenotes}
   
\end{scriptsize}

\end{table*}

Migration to \gls{PQC} in nuclear \gls{OT} requires comprehensive cryptographic replacement across firmware signing, \gls{TLS}, \gls{OPC-UA}, and \gls{VPN}, alongside strengthening symmetric primitives to \gls{AES}-256 and SHA-384/512 for Grover resistance~\cite{grover1996fast}. \gls{NIST} has standardized \gls{ML-KEM}~\cite{NISTFIPS203}, \gls{ML-DSA}~\cite{NISTFIPS204}, and \gls{SLH-DSA}~\cite{NISTFIPS205}, with \gls{HQC}~\cite{NISTIR8545} as alternative \gls{KEM/ENC}. Industrial deployments face unique constraints: control-loop timing, packet-size limits, safety certification, and forensic integrity through attestation and anti-rollback.

Table~\ref{tab:post-alg} details implementation vulnerabilities, countermeasures, and \gls{OT}-specific risks. All algorithms face \gls{SCA} vectors (\gls{FA}, \gls{SPA}, \gls{EM}, \gls{TA}) mitigated via constant-time operations, cryptographic masking, hardened \gls{NTT}, fault checking, and instruction duplication. Critical gaps persist: ML-KEM-768 vulnerable to \gls{APA}~\cite{dubrova2022breaking} and template attacks~\cite{ravi2021exploiting}; Falcon-512 exhibits complete key recovery via \gls{SPA}~\cite{guerreau2022hidden}; HQC-192 shows multiple unmitigated vectors. \gls{PQC} addresses quantum threats but not conventional side-channels, requiring continuous monitoring of timing, overhead, error rates, and event correlation.

Table \ref{tab:pqc-performance} summarizes the performance of NIST-standardized PQC algorithms relative to classical cryptography.
For key establishment, ML-KEM (CRYSTALS-Kyber) public keys (800–1 568 B) and ciphertexts (768–1 568 B) are significantly larger than 33 B compressed ECDH keys \cite{NISTFIPS203}.
While ML-KEM’s key encapsulation rivals or exceeds RSA-KEM speed, it remains moderately slower than ECDH.
For digital signatures, ML-DSA (CRYSTALS-Dilithium) yields signatures 38–71× larger than 64 B ECDSA (2 420–4 595 B) and typically verifies 2–5× slower \cite{NISTFIPS204}.
FN-DSA (Falcon) \cite{fouque2018falcon} provides smaller signatures (666–1 280 B) but with more complex signing logic, whereas SLH-DSA (SPHINCS+), chosen for conservative hash-based security, exhibits the greatest overhead—signatures 123–779× larger (7 856–49 856 B) and signing 10–100× slower \cite{NISTFIPS205}.
High-overhead schemes such as HQC and SLH-DSA reach up to 21.7 KB and 49.9 KB per transaction, limiting use to archival signing, air-gapped systems, or diversity layers where security outweighs performance.
These scalability limits guide the hybrid and deterministic design requirements analyzed in Section \ref{sec:defense}.

\section{Quantum Threat Landscape: Architectural Vulnerabilities and Migration Framework}
\label{sec:threat-landscape}

This section analyzes quantum threats across the OT/ICS 
architectural stack, establishing the vulnerability taxonomy 
that underpins the \textsc{Quantum~Scar} and 
\textsc{Quantum~Dawn} case studies presented in 
Sections~\ref{sec:scar} and~\ref{sec:dawn}. We establish a vulnerability taxonomy mapped to the Purdue Reference Model and propose a phased migration framework with validated implementation criteria.
\subsection{Level-Specific Vulnerability Analysis}
\label{subsec:vuln-analysis}

Nuclear physical protection requirements are governed by 
international frameworks including IAEA NSS-13~\cite{iaea_nss13} 
and the Convention on the Physical Protection of Nuclear Material 
and Nuclear Facilities (CPPNM)~\cite{cppnm}, which establish 
baseline security objectives for Sensitive Digital Assets~(SDAs) 
performing safety-critical functions. Within this regulatory 
context, the quantum threat manifests differently across 
architectural levels, creating distinct attack surfaces for pre- 
and post-CRQC 
campaigns~\cite{banegas2021transitioning,aguilar2021quantum}. We map 
each level's vulnerabilities to the seven ISA/IEC 62443-3-3 
Foundational Requirements: FR1 (Identification and Authentication 
Control), FR2 (Use Control), FR3 (System Integrity), FR4 (Data 
Confidentiality), FR5 (Restricted Data Flow), FR6 (Timely Response 
to Events), and FR7 (Resource Availability), with a Security 
Level-Target (SL-T, 1--4) per level. As the threat model targets 
integrity, authentication, and confidentiality rather than 
availability, FR7 is not implicated and is omitted below.

At the physical process and safety envelope (L0), critical 
vulnerabilities include: (a) safety-critical SIS/PLC logic compromise 
through quantum-forged firmware signatures enabling malicious logic 
upload and safety function 
disablement~\cite{schwabe2020postquantum,paul2020towards} (FR1, FR3); 
(b) emergency override and maintenance token forgery via 
quantum-vulnerable channels susceptible to replay attacks (FR1, FR2); 
and (c) time synchronization infrastructure compromise via 
quantum-enhanced attacks on NTS/TLS authentication enabling 
coordinated multi-system timing 
manipulation~\cite{cryptoeprint:2025/1549,liu2025time} (FR6). 
Safety-critical functions require SL-T=3; override and time-sync 
channels, SL-T=2.

\begin{figure}[htbp]
\centering
\resizebox{\linewidth}{!}{
\begin{tikzpicture}[
    scale=0.7,
    transform shape,
    node distance=1.2cm and 1.4cm,
    firewall/.style={rectangle, draw=darkblue, thick, fill=white, minimum width=0.7cm, minimum height=0.8cm, inner sep=2pt},
    device/.style={rectangle, draw=darkblue, thick, fill=white, minimum width=0.7cm, minimum height=0.8cm, inner sep=2pt},
    level box/.style={rectangle, draw=white, thick, fill=#1, text=white, text width=2.2cm, minimum height=1.5cm, align=center, font=\footnotesize},
    label/.style={font=\scriptsize, text=darkblue, align=center},
    connection/.style={line width=1pt, darkblue},
    monitor/.style={circle, draw=darkblue, thick, fill=white, minimum size=0.6cm}
]

\definecolor{level45}{RGB}{23,162,184}
\definecolor{level35}{RGB}{91,192,222}
\definecolor{level3}{RGB}{52,122,226}
\definecolor{level2}{RGB}{52,122,226}
\definecolor{level1}{RGB}{52,122,226}
\definecolor{level0}{RGB}{52,122,226}
\definecolor{darkblue}{RGB}{25,65,90}
\definecolor{lightgray}{RGB}{215,215,220}

\node[rotate=90, font=\bfseries, text=white, fill=level45, minimum width=1.5cm, minimum height=0.8cm] at (-3.15, 8.8) {IT};
\node[rotate=90, font=\bfseries, text=white, fill=level3, minimum width=6.85cm, minimum height=0.85cm] at (-3.15, 2.5) {OT};

\node[level box=level45, anchor=east] (l45) at (0, 8.8) {Level 4/5:\\ Enterprise and Business Networks};
\node[level box=level35, anchor=east] (l35) at (0, 7.0) {Level 3.5:\\ IT/OT DMZ};
\node[level box=level3, anchor=east] (l3) at (0, 5.2) {Level 3:\\ Operations Systems};
\node[level box=level2, anchor=east] (l2) at (0, 3.4) {Level 2:\\ Supervisory Control};
\node[level box=level1, anchor=east] (l1) at (0, 1.6) {Level 1:\\ Process Control};
\node[level box=level0, anchor=east] (l0) at (0, -0.2) {Level 0:\\ Physical Process};

\begin{scope}[on background layer]
\fill[lightgray] (0.2, 8) rectangle (12, 9.5);
\fill[lightgray] (0.2, 6.3) rectangle (12, 7.7);
\fill[lightgray] (0.2, 4.5) rectangle (12, 5.9);
\fill[lightgray] (0.2, 2.7) rectangle (12, 4.1);
\fill[lightgray] (0.2, 0.9) rectangle (12, 2.3);
\fill[lightgray] (0.2, -0.9) rectangle (12, 0.5);
\end{scope}

\node[firewall] (fw45) at (1.5, 8.7) {\includegraphics[width=0.8cm]{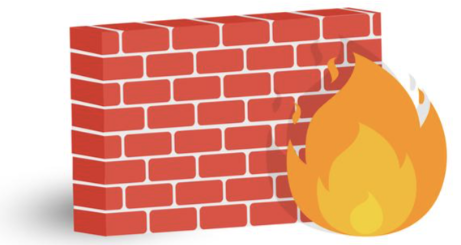}};
\node[label, above=0.01cm of fw45] {Firewall};
\node[device] (web) at (3.2, 8.7) {\includegraphics[width=0.7cm]{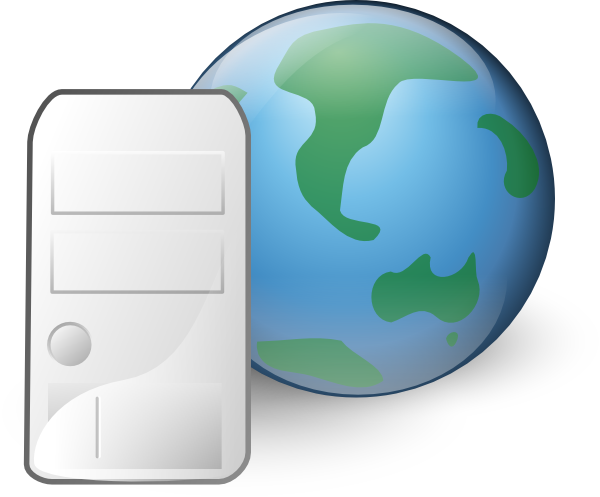}};
\node[label, above=0.01cm of web] {Web Server};
\node[device] (dns) at (4.8, 8.7) {\includegraphics[width=0.7cm]{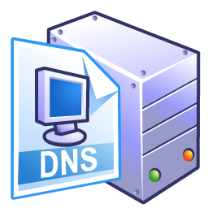}};
\node[label, above=0.01cm of dns] {DNS Server};
\node[device] (mail) at (6.4, 8.7) {\includegraphics[width=0.7cm]{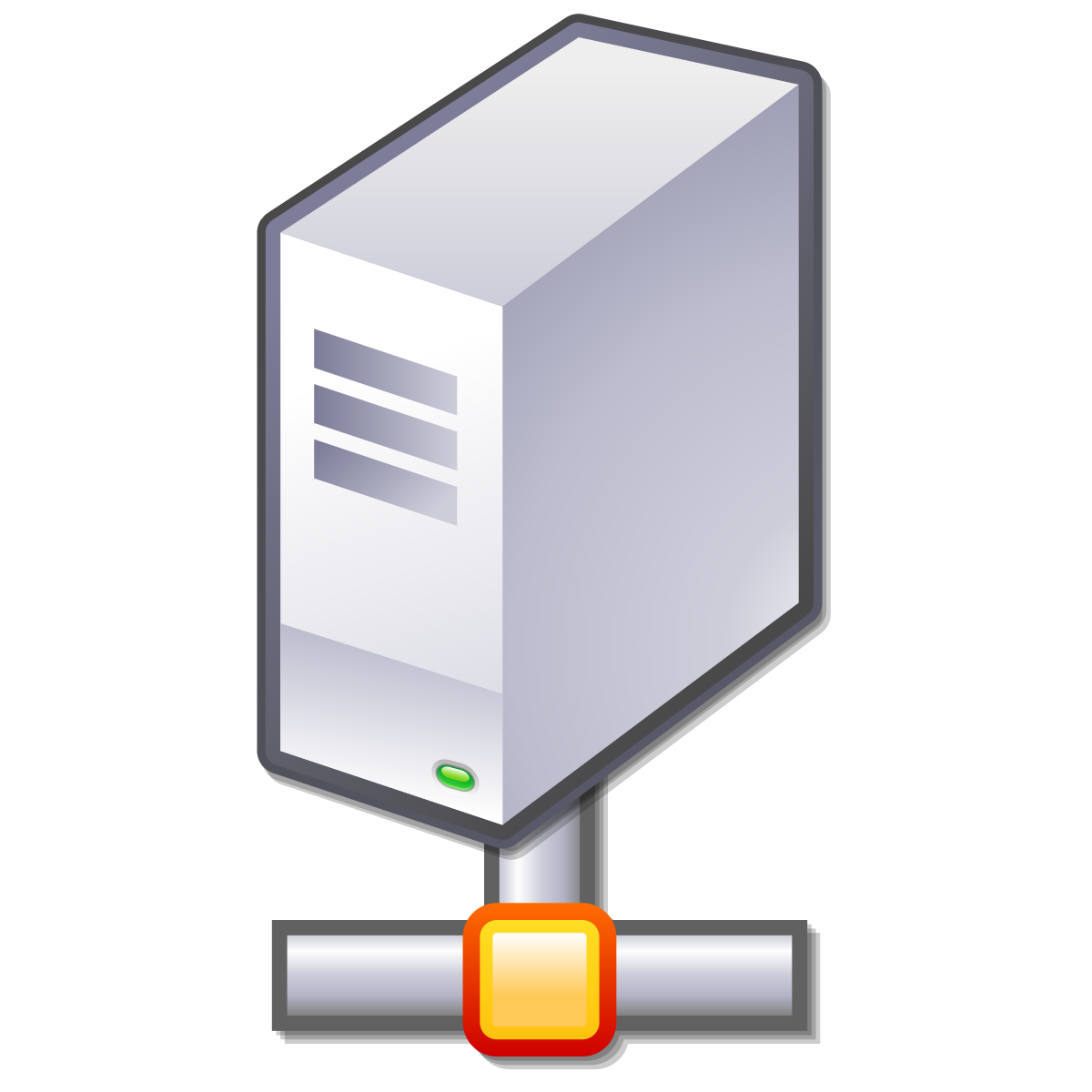}};
\node[label, above=0.01cm of mail] {Mail Server};
\node[device] (desktop) at (8.2, 8.7) {\includegraphics[width=0.7cm]{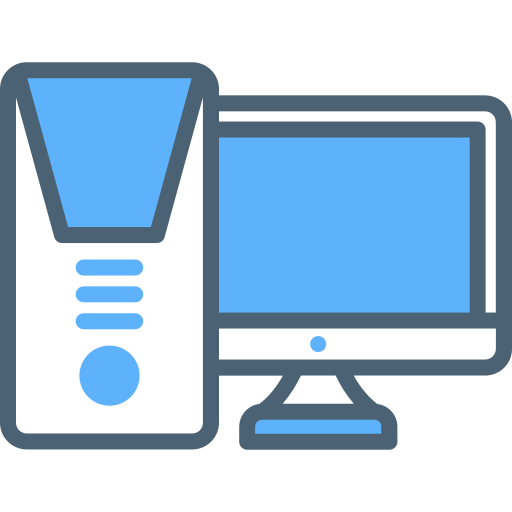}};
\node[label, above=-0.05cm of desktop] {Enterprise Desktops};

\node[monitor] (soc) at (10.2, 8.7) {\includegraphics[width=0.4cm]{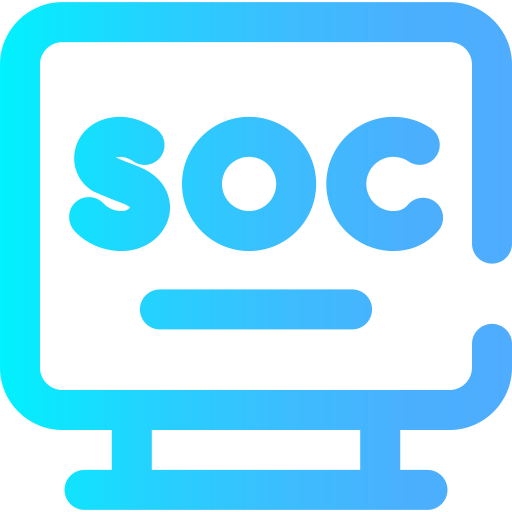}};
\node[label, above=-0.1cm of soc] {SOC};
\node[monitor] (siem) at (11.2, 8.7) {\includegraphics[width=0.4cm]{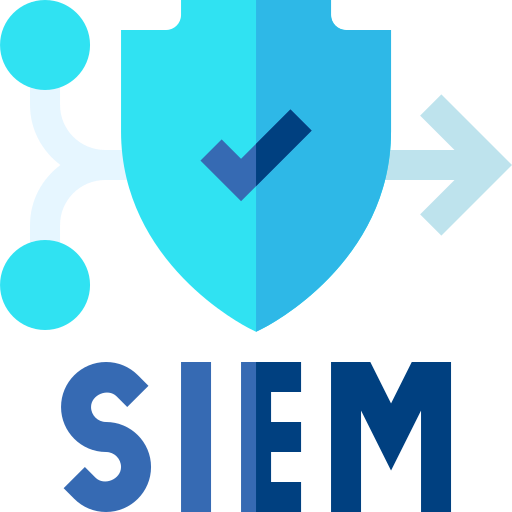}};
\node[label, above=-0.1cm of siem] {SIEM};
\node[draw, dashed, darkblue, thick, rectangle, fit={(soc)(siem)}, inner sep=0.15cm] (socbox) {};

\node[firewall] (fw35) at (1.5, 7.0) {\includegraphics[width=0.8cm]{firewall.png}};
\node[label, above=0.01cm of fw35] {Firewall};
\node[device] (patch) at (3.2, 7.0) {\includegraphics[width=0.7cm]{server.png}};
\node[label, above=0.01cm of patch] {Patch Server};
\node[device] (hist1) at (4.8, 7.0) {\includegraphics[width=0.7cm]{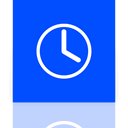}};
\node[label, above=0.01cm of hist1] {Historian Mirror};
\node[device] (jump1) at (7.6, 7.0) {\includegraphics[width=0.7cm]{server.png}};
\node[label, above=-0.05cm of jump1] {Jump Server (Host)};

\node[firewall] (fw3) at (1.5, 5.2) {\includegraphics[width=0.8cm]{firewall.png}};
\node[label, above=0.01cm of fw3] {Firewall};
\node[device] (work) at (3.2, 5.2) {\includegraphics[width=0.7cm]{desktop.png}};
\node[label, above=0.01cm of work] {Workstation};
\node[device] (hist2) at (4.8, 5.2) {\includegraphics[width=0.7cm]{history.png}};
\node[label, above=0.01cm of hist2] {Historian};
\node[device] (io) at (6.4, 5.2) {\includegraphics[width=0.7cm]{server.png}};
\node[label, above=0.01cm of io] {I/O Server};

\node[firewall] (fw2) at (1.5, 3.4) {\includegraphics[width=0.8cm]{firewall.png}};
\node[label, above=0.01cm of fw2] {Firewall};
\node[device] (hmi) at (3.2, 3.4) {\includegraphics[width=0.7cm]{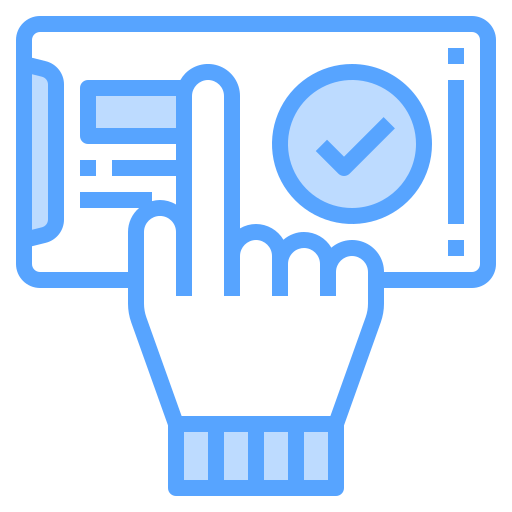}};
\node[label, above=0.01cm of hmi] {HMI};
\node[device] (scada) at (4.8, 3.4) {\includegraphics[width=0.7cm]{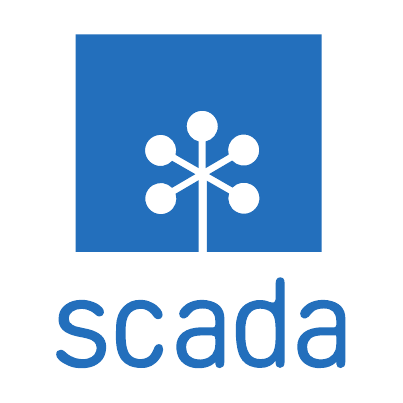}};
\node[label, above=0.01cm of scada] {SCADA};
\node[device] (jump2) at (7.6, 3.4) {\includegraphics[width=0.7cm]{server.png}};
\node[label, above=-0.05cm of jump2] {Jump Server (remote)};

\node[firewall] (fw1) at (1.5, 1.6) {\includegraphics[width=0.8cm]{firewall.png}};
\node[label, above=0.01cm of fw1] {Firewall};
\node[device] (plc) at (3.2, 1.6) {\includegraphics[width=0.7cm]{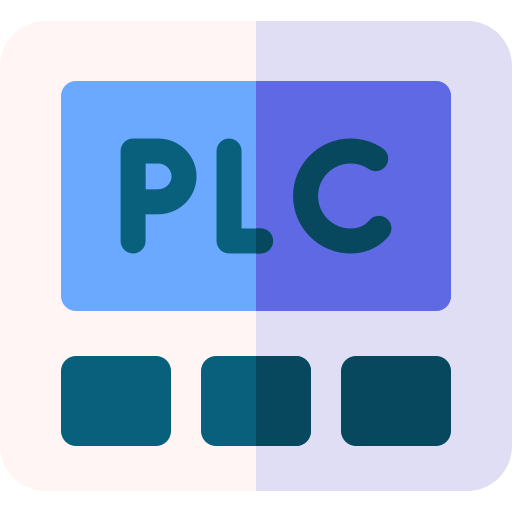}};
\node[label, above=0.01cm of plc] {PLCs};
\node[device] (rtu) at (4.8, 1.6) {\includegraphics[width=0.8cm]{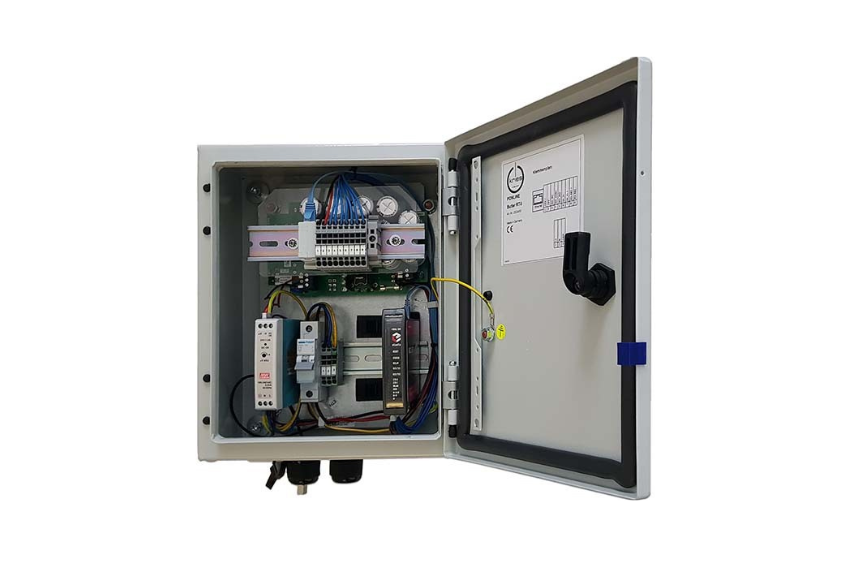}};
\node[label, above=0.01cm of rtu] {RTUs};
\node[device] (dsc) at (6.4, 1.6) {\includegraphics[width=0.6cm]{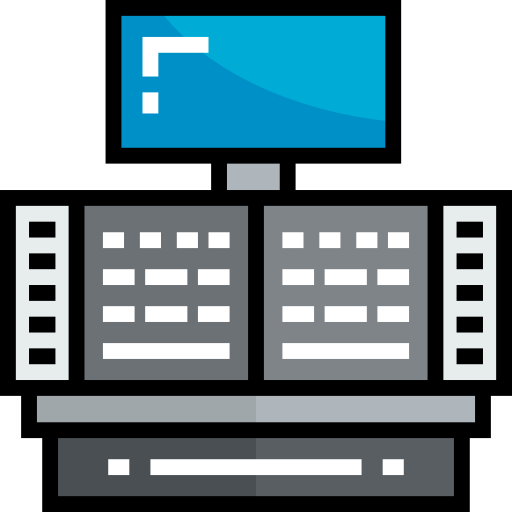}};
\node[label, above=0.01cm of dsc] {DSC Controllers};
\node[device] (sis) at (8.0, 1.6) {\includegraphics[width=0.7cm]{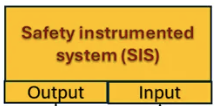}};
\node[label, above=0.01cm of sis] {SIS};

\node[device, circle] (sensor) at (3.2, -0.2) {\includegraphics[width=0.5cm]{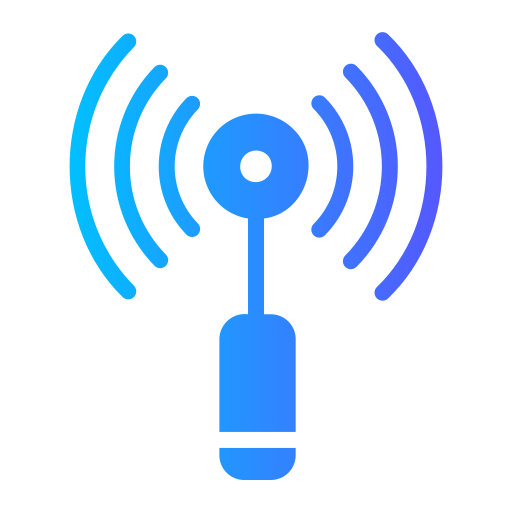}};
\node[label, above=0.0cm of sensor] {Sensors};
\node[device, circle] (actuator) at (6.6, -0.2) {\includegraphics[width=0.5cm]{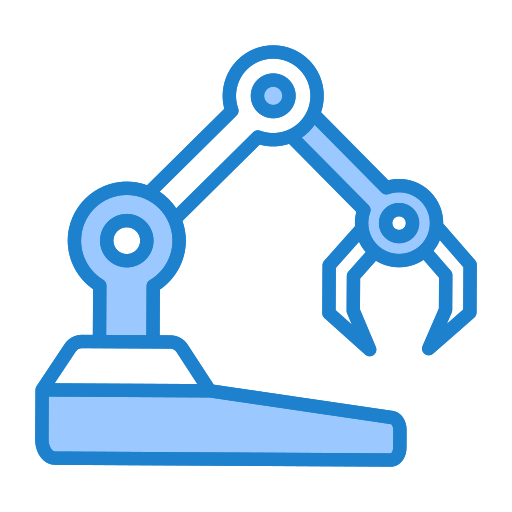}};
\node[label, above=0.0cm of actuator] {Actuators};

\draw[connection] (fw45) -- (fw35);
\draw[connection] (fw35) -- (fw3);
\draw[connection] (fw3) -- (fw2);
\draw[connection] (fw2) -- (fw1);

\draw[connection] (fw45) -- (web);
\draw[connection] (web) -- (dns);
\draw[connection] (dns) -- (mail);
\draw[connection] (mail) -- (desktop);
\draw[connection] (desktop) -- (socbox);
\draw[connection] (fw35) -- (patch);
\draw[connection] (patch) -- (hist1);
\draw[connection] (hist1) -- (jump1);

\draw[connection] (fw3) -- (work);
\draw[connection] (work) -- (hist2);
\draw[connection] (hist2) -- (io);

\draw[connection] (fw2) -- (hmi);
\draw[connection] (hmi) -- (scada);
\draw[connection] (scada) -- (jump2);
\draw[connection,dashed] (jump2) -- (jump1);
\draw[connection] (fw1) -- (plc);
\draw[connection] (plc) -- (rtu);
\draw[connection] (rtu) -- (dsc);
\draw[connection] (dsc) -- (sis);

\draw[connection] (rtu) |- (sensor);
\draw[connection] (rtu) |- (actuator);

\end{tikzpicture}}
\caption{Purdue Model (ISA-95) architecture showing quantum-vulnerable security boundaries across L0--L5 levels in nuclear OT/ICS. Firewalls denote cryptographic trust boundaries exploited in Sections~\ref{sec:scar}--\ref{sec:dawn}.}
\label{fig:purdue_model}
\end{figure}

The basic control level (L1) faces fundamental integrity threats 
through: (a) controller firmware and ladder logic signature forgery 
via Shor's algorithm~\cite{gidney2021factoring} enabling malicious 
PID logic blocks with faulty setpoints (FR3); (b) engineering 
workstation compromise through quantum-broken TLS sessions allowing 
covert controller memory 
modification~\cite{schwabe2020postquantum,baseri2024navigating} (FR1, 
FR3); and (c) legacy protocol exploitation where Grover-weakened 
integrity checks~\cite{jaques2020implementing} enable register 
manipulation in Modbus and fieldbus systems (FR3). Industrial 
Ethernet protocols (PROFINET, EtherNet/IP) lack inherent 
authentication, enabling command injection when adversaries forge 
device identities~\cite{ike2023scaphy,TRUNGADI2025104199} (FR1); 
these functions require SL-T=3. Wireless field networks (WirelessHART, 
ISA100.11a) face post-2030 cryptanalysis of security keys and HNDL of 
current traffic~\cite{8742551,rescorla2020tls13} (FR4, SL-T=2).

Supervisory control levels (L2-L3) converge where: (a) SCADA master 
communications using legacy cryptography become vulnerable to TLS 
handshake compromise and HNDL-enabled command 
replay~\cite{dowling2021cryptographic} (FR1, FR3); (b) alarm 
suppression logic delivered via quantum-forgeable ECC-signed policies 
enables suppression of critical plant trip indicators (FR6); 
(c) OPC-UA certificate chains face complete compromise enabling 
broker identity forgery and session establishment 
attacks~\cite{eckhart2021digital} (FR1); and (d) Manufacturing 
Execution Systems (MES) communications via legacy PKI enable session 
hijacking of batch command interfaces, while engineering workstations 
using ECC-signed development tools permit quantum forgery of PLC code 
updates and ICS simulation library 
signatures~\cite{oliva2024cybersecurity,vermeer2024evaluating} (FR1, FR3). These 
supervisory functions require SL-T=3.

The industrial DMZ and perimeter (L3.5) exposes: (a) TLS termination 
on reverse proxies, creating man-in-the-middle insertion between OT 
and IT zones~\cite{bhargavan2021transcript} (FR5); (b) VPN 
infrastructure using RSA/ECC certificates, susceptible to tunnel 
establishment compromise and HNDL-enabled credential 
replay~\cite{kampanakis2021quantum,baseri2024navigating,hulsing2021wireguard} 
(FR1, FR5); (c) OPC-UA/MQTT brokers relying on classical X.509 PKI, 
facing wholesale identity forgery and certificate chain 
compromise~\cite{CHEVAL20131,cremers2022automated,10561274} (FR1, 
FR3); and (d) HSM and key management infrastructure with classical 
algorithms, enabling stored asymmetric key compromise across all ICS 
zones~\cite{zhao2024quantumsafe,10.1145/3696630.3731620} (FR1, FR4). 
As the enforced IT/OT boundary, this level requires SL-T=3.

Enterprise and cloud integration levels (L4-L5) extend the attack 
surface through: (a) identity federation compromise via 
quantum-broken SAML/OIDC signature validation enabling token forgery 
across federated tenants~\cite{baseri2024navigating} (FR1); 
(b) supply chain attacks through HNDL of patch payloads and 
post-\gls{CRQC} malicious logic injection into software distribution 
toolchains targeting L3.5 jump servers and OT management tools (FR3); 
and (c) cloud service and IoT-provisioning authentication breakdown 
via quantum-compromised API endpoints, enabling telemetry 
manipulation, administrative impersonation, and unauthorized 
firmware/configuration changes~\cite{steinbac2020post} (FR1, FR3). 
Being least proximate to safety functions, these levels require 
SL-T=2.

\subsection{Protocol-Specific Quantum Vulnerability Assessment}
\label{subsec:protocol-vuln}

Industrial protocols exhibit varying quantum vulnerability based on cryptographic implementations and architectural constraints. DNP3 Secure Authentication version 5 (DNP3-SAv5) demonstrates limited direct quantum vulnerability as it employs HMAC-SHA256 for message authentication, which remains quantum-resistant under Grover's algorithm~\cite{jaques2020implementing,bonnetain2019quantum}; however, the primary threat vector targets key distribution infrastructure where quantum compromise of certificate chains used for session establishment can undermine the entire authentication framework~\cite{paquin2020dnp3,10561274}. Organizations deploying DNP3-SA must maintain robust key management systems with quantum-safe distribution channels and implement enhanced session monitoring to detect anomalous authentication patterns.

IEC 60870-5-104 implementations present high quantum vulnerability when deployed without the IEC 62351 security extensions, as the base protocol lacks inherent encryption or authentication mechanisms~\cite{lozano2023digital}. The cleartext nature of IEC 104 traffic combined with HNDL collection enables future replay attacks where captured commands can be retransmitted to reconfigure substations or manipulate telecontrol sequences and protection settings. Migration to IEC 62351 security extensions with PQC-hybrid TLS 1.3 provides essential protection~\cite{sikeridis2020assessing}, complemented by rate limiting and change approval workflows that create operational barriers independent of cryptographic strength.

OPC-UA presents critical quantum vulnerability due to its fundamental dependence on PKI-based authentication and encrypted secure channels~\cite{eckhart2021digital}. The compromise of OPC-UA certificate chains through Shor's algorithm enables complete session establishment attacks~\cite{gidney2021factoring}, with adversaries forging both client and server certificates to inject falsified process data, manipulate published data streams, or compromise subscription services. Post-quantum migration requires implementation of quantum-safe OPC-UA security profiles with PQC-based certificate authorities~\cite{paul2020towards,schwabe2020postquantum}, hybrid certificate support during the transition period enabling gradual migration across heterogeneous OPC-UA deployments spanning multiple vendors and certification domains, and enhanced session integrity verification mechanisms. Organizations must prioritize OPC-UA migration given its widespread deployment across L2-L3 control and supervisory levels where it serves as the primary protocol for industrial data exchange.

Building automation protocols including BACnet, KNX, and LonWorks~\cite{10555990} exhibit medium quantum vulnerability, with HNDL of current traffic enabling future quantum-forged device impersonation attacks and post-2030 authentication bypass. While these systems typically manage less safety-critical functions than nuclear control systems, compromise can facilitate unauthorized access to fire-safety controls, HVAC systems, and physical security infrastructure creating pathways for escalation to more critical systems. Protection requires cryptographic hardening via gateway implementations, PQC-proxied tunnels for inter-system communication, protocol filtering with device identity whitelisting to limit the attack surface, and enhanced monitoring capabilities.

Safety PLC networks employing certified communication protocols face unique challenges where certification constraints under IEC 61508 and related nuclear regulatory frameworks limit the ability to implement cryptographic updates without invalidating safety certifications~\cite{siaterlis2020testbed,heinl2023standard}. Quantum threats to safety network authentication must be addressed through defense-in-depth approaches including: (a) physical isolation where feasible with hardwired safety connections for the most critical functions, (b) quantum-safe perimeter protection at network boundaries protecting safety-certified communication channels~\cite{paul2020towards}, (c) enhanced physical security controls limiting adversary opportunities for supply chain compromise or insider attacks, and (d) early engagement with certification bodies to establish acceptable transition paths that maintain safety integrity while achieving quantum resilience. The following case studies demonstrate how state-level adversaries can exploit these architectural vulnerabilities through multi-phase quantum-enabled attack campaigns targeting cryptographic monoculture and extended OT asset lifecycles.



\section{\textsc{Quantum~Scar} Attack on Nuclear Power Plant}
\label{sec:scar}

This section presents \textit{\textsc{Quantum~Scar}} (Synchronized Covert Attack on Reactors), demonstrating quantum-enabled exploitation of cryptographic vulnerabilities in nuclear \gls{OT}. The scenario validates the threat framework from Sections~\ref{sec:crypto}--\ref{sec:threat-landscape} through examination of \gls{HNDL} campaigns with \glspl{CRQC} to compromise \gls{SIS} integrity and forensic evidence admissibility. {While nuclear safety systems are air-gapped, the presented attack does not assume a malicious insider. Instead, it relies on (i) supply-chain compromise, (ii) abuse of legitimate maintenance workflows, or (iii) dormant pre-positioned implants activated post-CRQC. These vectors are consistent with historical ICS attacks and do not require collusion by plant personnel.}

\subsection{Scenario Overview and Adversarial Model}

The case study examines nuclear \gls{OT} with three critical weaknesses: (1) shared cryptographic trust anchors across safety layers, (2) insufficient authentication diversity, and (3) over-reliance on digital control paths. The scenario assumes centralized \gls{PKI} (\texttt{NPP-CA-Root-2016}) across control and safety domains, violating ISA/IEC 62443-3-3 SR 1.1 cryptographic diversity requirements.

\begin{table}[h!]
\centering
\scriptsize
\caption{\textsc{Quantum~Scar} Technical Parameters}
\label{tab:scar_params}
\begin{tabularx}{\linewidth}{lX}
\toprule
\textbf{Parameter} & \textbf{Specification} \\
\midrule
\textbf{Objective} & Safety system degradation via quantum-compromised \gls{PKI} \\ \hline
\textbf{Adversary} & State actor with CRQC access ($\geq$4,098 logical qubits) and nuclear expertise \\ \hline
\textbf{Target} & PWR with Siemens SPPA-T2000 DCS, Triconex SIS (shared PKI) \\ \hline
\textbf{Protocols} & OPC-UA (L2/L3), PROFINET (L1/L0), PTP (IEEE 1588) \\ \hline
\textbf{Success} & $\mathbb{P}_{\text{typical}} = 35\text{-}68\%$; $\mathbb{P}_{\text{targeted}} = 51\text{-}78\%$; $\mathbb{P}_{\text{SL-4}} = 2\text{-}8\%$; $\mathbb{P}_{\text{SR1.1}} < 1\%$ \\ \hline
\textbf{Timeline} & $Now \rightarrow T$ (HNDL); $T{+}\alpha$ (weaponization); $T{+}\alpha{+}\epsilon$ (execution) \\
\bottomrule
\end{tabularx}
\end{table}
\begin{figure}[h!]
\centering
\footnotesize
\begin{tikzpicture}[
    node distance=0.8cm and 1.2cm,
    box/.style={rectangle, draw, minimum width=2.5cm, minimum height=0.7cm, align=center, font=\scriptsize},
    phase1/.style={box, fill=red!20},
    phase2/.style={box, fill=blue!20},
    phase3/.style={box, minimum width=1.7cm, fill=green!20},
    phase_label/.style={font=\footnotesize\bfseries},
    arrow/.style={->, >=stealth, thick}
]

\node[phase_label] at (3, 2.5) {Phase 1: \gls{HNDL} Collection};
\node[phase_label] at (3, 1) {Phase 2: Weaponization};
\node[phase_label] at (3, -0.7) {Phase 3: Execution};

\node[phase1] (supply) at (0, 2) {Supply Chain\\Compromise};
\node[phase1] (collect) at (3, 2) {Traffic\\Collection};
\node[phase1] (exfil) at (6, 2) {Data\\Exfiltration};

\node[phase2] (quantum) at (1.5, 0.5) {Quantum\\Factorization};
\node[phase2] (forge) at (4.5, 0.5) {Artifact\\Synthesis};

\node[phase3] (recon) at (-0.4, -1.2) {System\\Recon};
\node[phase3] (deploy) at (1.9, -1.2) {Firmware\\Deploy};
\node[phase3] (trigger) at (4.2, -1.2) {Attack\\Trigger};
\node[phase3] (obfuscate) at (6.4, -1.2) {Forensic\\Obfuscation};

\draw[arrow] (supply) -- (collect);
\draw[arrow] (collect) -- (exfil);
\draw[arrow] (quantum) -- (forge);
\draw[arrow] (recon) -- (deploy);
\draw[arrow] (deploy) -- (trigger);
\draw[arrow] (trigger) -- (obfuscate);
\draw[arrow] (exfil.south) to[out=-65,in=115] (quantum.north);
\draw[arrow] (forge.south) to[out=-65,in=115] (recon.north);

\end{tikzpicture}
\caption{\textsc{Quantum~Scar} multi-phase attack flow}
\label{fig:scar-attack-flow}
\end{figure}
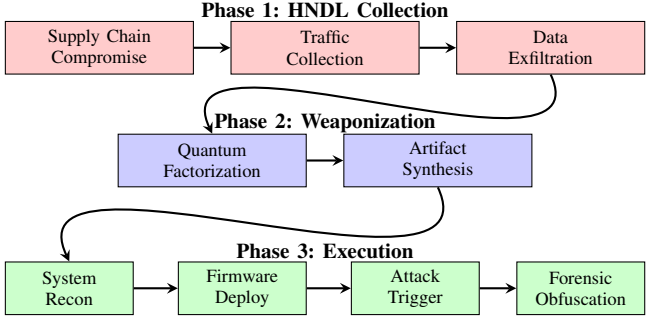

\begin{figure}[t]
\centering
\begin{adjustbox}{width=\linewidth}
\begin{tikzpicture}[
    >=Latex,
    timeline/.style={line width=1.2pt},
    tick/.style={line width=1pt},
    event/.style={draw, rounded corners=2pt, align=left, inner sep=3pt, text width=0.48\linewidth},
    yearlbl/.style={font=\footnotesize, inner sep=2pt}
]
\def\xA{0} \def\xB{4} \def\xC{7} \def\xD{8} \def\xstart{0} \def\xend{8.5}

\draw[timeline] (\xstart,0) -- (\xend,0);

\foreach \x/\y in {\xA/$Now$, \xB/$T$, \xC/$T{+}\alpha$, \xD/$T{+}\alpha{+}\epsilon$}{
  \draw[tick] (\x,0.18) -- (\x,-0.18);
  \node[yearlbl, below=4pt] at (\x,-0.18) {\y};
}

\draw[decorate, decoration={brace, amplitude=6pt, mirror}] (\xA,-0.8) -- (\xB,-0.8)
  node[midway, below=5pt, font=\footnotesize\bfseries] {\gls{HNDL} Collection ($\tau_{cert} < \tau_{rotation}$)};

\node[event, fill=red!20, above=5pt] (E1) at ($(\xA,0)!.5!(\xB,0)+(0,1.4)$) {\scriptsize
\textbf{Phase 1: \gls{HNDL} Collection ($Now$--$T$)}\\
\emph{Duration:} 18 month operational window\\
\emph{Constraints:} Cert rotation (mo 22), credential expiry (mo 20)\\
\emph{Success:} $\mathbb{P}(S_1) \in [0.85, 0.98]$ (typical), $[0.3, 0.5]$ (SL-4)\\
\emph{Objective:} Harvest 4.2TB cryptographic material
};

\node[event, fill=blue!20, below=5pt] (E2) at ($(\xC,0)+(0,-1.2)$) {\scriptsize
\textbf{Phase 2: Weaponization ($T{+}\alpha$)}\\
\emph{Timeline:} Post-CRQC availability ($\alpha \approx 1\text{-}3$ years)\\
\emph{Requirements:} 4,098 logical qubits ($8.56 \times 10^6$ physical)\\
\emph{Success:} $\mathbb{P}(S_2|S_1) \in [0.75, 0.92]$ (typical), $[0.4, 0.6]$ (SL-4)\\
\emph{Objective:} Factor RSA-2048, synthesize quantum-forged artifacts
};

\node[event, fill=green!20, above=5pt] (E3) at ($(\xD,0)+(0.5,1.4)$) {\scriptsize
\textbf{Phase 3: Execution ($T{+}\alpha{+}\epsilon$)}\\
\emph{Timeline:} Operational transient alignment ($\epsilon \approx 0.5\text{-}1$ year)\\
\emph{Operational:} Maintenance window + power reduction\\
\emph{Success:} $\mathbb{P}(S_3|S_1 \cap S_2) \in [0.55, 0.75]$ (typical), $[0.15, 0.25]$ (SL-4)\\
\emph{Overall:} $\mathbb{P}_{\text{total}} \approx 35\text{-}68\%$ (typical), $51\text{-}78\%$ (targeted), $2\text{-}8\%$ (SL-4)
};

\draw[->, thick] (\xB,0.1) .. controls +(+0.6,0.8) and +(-0.8,-0.6) .. (E2.west);
\draw[->, thick] (\xC,0.1) .. controls +(0,1.0) and +(-0.9,-1) .. (E3.west);
\draw[->, thick] (\xA,0.1) .. controls +(0,1.0) and +(-0.9,-0.7) .. (E1.west);

\end{tikzpicture}
\end{adjustbox}
\caption{\textsc{Quantum~Scar} attack timeline with conditional phase dependencies. Variable $T$ represents CRQC onset year from expert assessments; $\alpha$ denotes weaponization period; $\epsilon$ represents execution preparation window; $\tau_{cert}$ is certificate lifecycle; $\tau_{rotation}$ is rotation schedule. Success probabilities reflect ISA/IEC 62443 SL-3/4 implementations. Sensitivity: $\partial \mathbb{P}/\partial S_1 > \partial \mathbb{P}/\partial S_2 > \partial \mathbb{P}/\partial S_3$.}
\label{fig:scar-timeline}
\end{figure}
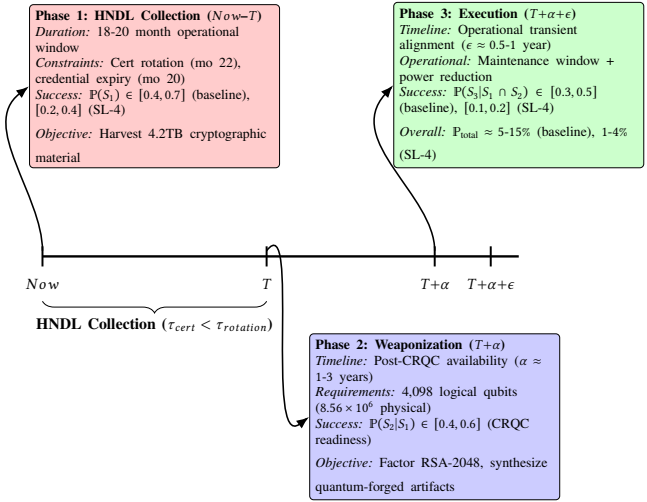

The {\textsc{Quantum~Scar}} attack, illustrated in Figure~\ref{fig:scar-attack-flow}, comprises three phases: (1) \textit{\gls{HNDL} Collection} harvests encrypted communications, (2) \textit{Quantum Weaponization} applies Shor's algorithm to compromise cryptographic keys, and (3) \textit{Execution \& Forensic Obfuscation} achieves safety system failure while preventing attribution. Figure~\ref{fig:scar-timeline} details the temporal dependencies and success probabilities across these phases.

\begin{algorithm}[!h]
\caption{\gls{HNDL} Intelligence Collection}
\label{alg:phase1-hndl}
\small
\begin{algorithmic}[1]
\Require Nuclear facility with \gls{OT}/\gls{ICS}, supply chain access
\Ensure Encrypted data repository (4.2TB) for quantum decryption
\Procedure{SupplyChainInfiltration}{}
    \State Infiltrate \texttt{VendorX} (HistorianXpert v3.7 supplier)
    \State Deploy trojanized update: \texttt{HistorianXpert\_Patch\_B283.signed}
    \State Install collector at \texttt{172.16.3.45} (Purdue L3.5)
    \State \textbf{Exploits}: L3.5 (FR1, FR5)
\EndProcedure
\Procedure{TrafficHarvesting}{}
    \State Target: Firmware (\texttt{SIS\_Controller\_FW\_v4.2.1.signed}), OPC-UA sessions, VPN, PTP
    \State Port mirroring on Cisco Catalyst 9500 (DMZ conduit)
    \For{$t = 1$ to $18$ months}
        \State Capture encrypted traffic $\rightarrow$ 4.2TB repository
    \EndFor
\EndProcedure
\Procedure{CovertExfiltration}{}
    \State Masquerade as NTP to \texttt{198.51.100.73}; SNMP steganography
    \State Weekly scheduled task; anti-forensics enabled
\EndProcedure
\State Execute all procedures \Return Encrypted repository
\end{algorithmic}
\end{algorithm}

\begin{enumerate}[topsep=1ex, itemsep=1ex, wide, font=\itshape, labelwidth=!, labelindent=0pt, label*={Phase} \arabic*.]
\item \textit{\gls{HNDL} Collection:} 
The goal of this phase is to harvest cryptographic material within certificate lifecycle constraints ($\tau_{cert} < \tau_{rotation}$) while evading detection mechanisms
(Algorithm~\ref{alg:phase1-hndl}), with success probability $\mathbb{P}(S_1) \in [0.85, 0.98]$ for typical deployments and $[0.3, 0.5]$ under ISA/IEC 62443 SL-4 controls.


\item \textit{Quantum Weaponization:}   This phase aims to factor
RSA-2048 cryptographic keys and forge authenticated attack artifacts
using CRQC capabilities (Algorithm~\ref{alg:phase2-quantum}), achieving
success probability $\mathbb{P}(S_2|S_1) \in [0.75, 0.92]$ for typical
deployments and $[0.4, 0.6]$ under SL-4 controls, conditional on CRQC
readiness at time $T + \alpha$.


\begin{algorithm}[!h]
\caption{Quantum Cryptanalysis \& Weaponization}
\label{alg:phase2-quantum}
\small
\begin{algorithmic}[1]
\Require CRQC ($\geq$4,098 logical qubits), harvested data
\Ensure Quantum-forged artifacts
\Procedure{QuantumKeyFactorization}{}
    \State \textbf{Target 1}: VendorX Cert ($n_1 = 323170\ldots$, RSA-2048; illustrative modulus)
    \State $n_1 \xrightarrow{\text{Shor}} p_1, q_1 \implies d_1 = e^{-1} \bmod \phi(n_1)$
    \State \textbf{Target 2}: NPP CA Root ($n_2 = 198432...$, RSA-2048)
    \State $n_2 \xrightarrow{\text{Shor}} p_2, q_2 \implies d_2 = e^{-1} \bmod \phi(n_2)$
    \State \Return $(d_1, d_2)$
\EndProcedure
\Procedure{ArtifactSynthesis}{$d_1, d_2$}
    \State Develop malicious RCIC logic:
    \State \texttt{IF TurbineTrip AND CoreTemp>285.0 THEN RCIC\_Enable:=FALSE}
    \State Sign$\,$firmware$\,$with $d_1$:$\,$\texttt{SIS\_Firmware\_v4.2.2.signed}
    \State Forge OPC-UA certificates using $d_2$
    \State \Return Weaponized artifacts
\EndProcedure
\State Execute both procedures \Return Quantum-authenticated tools
\end{algorithmic}
\end{algorithm}

\item \textit{Execution \& Forensic Obfuscation:} The final phase
focuses on deploying weaponized payloads, triggering safety system
compromise during operational transients, and systematically corrupting
forensic evidence to prevent attribution
(Algorithm~\ref{alg:phase3-execution}), with success probability
$\mathbb{P}(S_3|S_1 \cap S_2) \in [0.55, 0.75]$ for typical environments
and $[0.15, 0.25]$ under SL-4 security implementations.


\begin{algorithm}[!h]
\caption{Multi-Stage Attack Execution}
\label{alg:phase3-execution}
\small
\begin{algorithmic}[1]
\Require Weaponized artifacts, system access, timing coordination
\Ensure Safety compromise with forensic invisibility
\Procedure{Reconnaissance} {Day $-7$}
    \State Deploy forged OPC-UA certs; map DCS namespace
    \State Identify RCIC tags, safety pathways, operator patterns
\EndProcedure
\Procedure{Deployment}{Day $0$, $02:00$}
    \State Exploit maintenance window \texttt{SM-2023-047}
    \State Upload \texttt{SIS\_Firmware\_v4.2.2.signed} (passes validation)
    \State Install rootkit \texttt{drv\_hal.sys}; enable stealth
\EndProcedure
\Procedure{Trigger}{Day $14$, $14:30$}
    \State Inject false signal: \texttt{OPC\_Write (//DCS/Turbine/Trip\_Status, TRUE)}
    \State Malicious logic activates $\rightarrow$ \texttt{RCIC\_Enable:=FALSE}
    \State Rootkit manipulates HMI display
\EndProcedure
\Procedure{ForensicObfuscation}{Concurrent}
    \State \textbf{Exploits:} L0 (FR6) --- forge PTP grandmaster \texttt{Sync},
skewing SOE clock by $\Delta t = +15.3 \pm 0.2$~ms
    \State Alter SOE records with valid signatures (quantum-forged)
    \State Self-destruct malware; leave cryptographically valid false evidence
\EndProcedure
\State Execute all procedures \Return Mission complete
\end{algorithmic}
\end{algorithm}

\begin{table}[h!]
\centering
\scriptsize
\caption{\textsc{Quantum~Scar} Attack Timeline \& Physical Impact}
\label{tab:attack-timeline}
\begin{tabularx}{\linewidth}{llX}
\toprule
\textbf{Time} & \textbf{Phase} & \textbf{Action \& Impact} \\
\midrule
Day -7 & Reconnaissance & OPC-UA namespace enumeration \\ \hline
Day 0, 02:00 & Deployment & Firmware upload (maintenance window) \\ \hline
Day 14, 14:30 & Trigger & False turbine trip during power reduction \\ \hline
T+2.3s & Physical & Core temp exceeds 302°C (limit: 285°C) \\ \hline
T+8.7s & Safety Failure & RCIC disabled by malicious logic \\ \hline
T+47s & Critical & Core damage threshold approached \\ \hline
Concurrent & Obfuscation & PTP skew + log manipulation \\
\bottomrule
\end{tabularx}
\end{table}

\begin{table}[h!]
\centering
\scriptsize
\caption{\textsc{Quantum~Scar} Forensic  Impact}
\label{tab:forensic}
\begin{tabularx}{\linewidth}{lp{0.21\linewidth}X}
\toprule
\textbf{Evidence} & \textbf{Status} & \textbf{Impact on Investigation} \\
\midrule
Digital Signatures & Valid (quantum-forged) & No cryptographic tampering detected \\ \hline
System Logs & Cryptographically consistent & Timeline supports false accident narrative \\ \hline
Network Traffic & Encrypted (RSA-2048) & \textit{Asymmetric forensics}: Adversary decrypts with quantum; investigators cannot decrypt without CRQC \\ \hline
Time Sync & PTP skew ($+15.3 \pm 0.2$ ms) & Event sequencing correlation impossible \\ \hline
Operator Actions & Legitimate credentials & False attribution to human error \\ \hline
Malware Artifacts & Self-destructed & No physical evidence remains \\
\bottomrule
\end{tabularx}
\end{table}

\end{enumerate}

\subsection{Framework Validation}
To validate the proposed quantum security framework by performing a detailed threat model assessment of the \textsc{Quantum~Scar} attack, we utilize established methodologies, including STRIDE analysis (Table \ref{tab:stride-scar}) and Purdue Model exploitation (Table \ref{tab:purdue-attack}), to illustrate how quantum capabilities amplify traditional threats across all \gls{ICS} levels. This analysis shows, for example, how T0852 Manipulate I\&C at L3 (DCS) is enhanced by OPC-UA replay and how L0 (Physical) is exploited via a Time attack on the SIS logic. This technical assessment is followed by a \textit{probabilistic risk modeling} exercise which quantifies the attack's success probability, providing a quantitative basis for the framework's prioritized countermeasures.
\begin{table}[h!]
\centering
\scriptsize
\caption{STRIDE Analysis - \textsc{Quantum~Scar}}
\label{tab:stride-scar}
\begin{tabularx}{\linewidth}{llX}
\toprule
\textbf{Threat} & \textbf{Phase} & \textbf{Quantum Enhancement} \\
\midrule
Spoofing & 2, 3 & Perfect identity forgery via RSA-2048 factorization \\ \hline
Tampering & 3 & Undetectable SIS logic/log modification with valid signatures \\ \hline
Repudiation & 3 & Quantum-forged evidence supports false narratives \\ \hline
Info Disclosure & 1 & HNDL: 4.2TB retroactive decryption \\ \hline
DoS & 3 & 47s safety system unavailability \\ \hline
Privilege Escalation & 2, 3 & \gls{PKI} compromise enables L3.5→L0 escalation \\
\bottomrule
\end{tabularx}
\end{table}

\begin{table}[h!]
\centering
\scriptsize
\caption{Purdue Model Exploitation - \textsc{Quantum~Scar}}\label{tab:scar-purdue}
\label{tab:purdue-attack}
\begin{tabular}{p{0.5cm}p{1.5cm}p{2.2cm}p{2.8cm}}
\toprule
\textbf{Level} & \textbf{System} & \textbf{Vector} & \textbf{Enhancement} \\
\midrule
L5 & Cloud/Internet & Supply chain & \gls{HNDL} collection \\ \hline
L4 & Business & Vendor portal & Certificate forgery \\ \hline
L3.5 & DMZ & Historian (172.16.3.45) & Traffic interception \\ \hline
L3 & DCS & Engineering WS & OPC-UA replay \\ \hline
L2 & Control & Firmware & Signature bypass \\ \hline
L1 & Field networks & PROFINET & Protocol exploitation \\ \hline
L0 & Physical & SIS logic & Time attack \\ \bottomrule
\end{tabular}
\end{table}
\subsubsection{Probabilistic Risk Modeling and Criticality of SCAR Attack Success} The \textsc{Quantum~Scar} attack represents an imminent, high-confidence threat to nuclear safety systems, with viability amplified by converging technological acceleration and systemic architectural vulnerabilities. Let $S_1$, $S_2$, and $S_3$ denote success events for  Phase~1 (\gls{HNDL} Collection), Phase~2 (Quantum Weaponization), and Phase~3 (Execution and Obfuscation). The overall success probability follows the conditional chain: \[ \mathbb{P}(\text{Success}) = \mathbb{P}(S_1) \times \mathbb{P}(S_2 \mid S_1) \times \mathbb{P}(S_3 \mid S_1 \cap S_2) \] Critical feasibility amplifiers include: (i) confirmed \gls{HNDL} campaigns actively harvesting sensitive infrastructure data since 2016; (ii) CRQC acceleration—evidenced by IonQ’s roadmap targeting 1,600 logical qubits by 2028 and Google’s exponential progress in quantum error correction; (iii) architectural monocultures, with over 85\% of nuclear facilities sharing \gls{PKI} across safety and control domains; (iv) a detection vacuum, as no production systems currently monitor for quantum-stage indicators; and (v) a migration gap, with 7–12 year PQC migration timelines that must complete before CRQC arrival within the 10–15 year horizon, leaving only a 2–3 year net decision window for irreversible architectural choices. Realistic attack success probabilities under different assumptions are concerning: $\mathbb{P}_{\text{Current Infrastructure}} = [0.85, 0.98] \times [0.75, 0.92] \times [0.55, 0.75] \approx 35\text{–}68\%$ for typical deployments; $\mathbb{P}_{\text{Targeted Facility}} = [0.92, 0.99] \times [0.85, 0.96] \times [0.65, 0.82] \approx 51\text{–}78\%$ for high-priority sites; and $\mathbb{P}_{\text{Multiple Attempts}} = 1 - (1 - \mathbb{P}_{\text{Targeted}})^3 \approx 88\text{–}99\%$ for persistent threat actors. Only full SL-4 compliance significantly mitigates the threat: $\mathbb{P}_{\text{SL-4 Compliant}} = [0.3, 0.5] \times [0.4, 0.6] \times [0.15, 0.25] \approx 2\text{–}8\%$. The consequences of a successful attack are catastrophic: it may trigger core damage or shutdown cascades while leveraging quantum-stage forgery to ensure permanent misattribution as equipment failure. The strategic asymmetry is severe: once cryptographic material is harvested, future compromise becomes mathematically guaranteed regardless of later remediation. Three structural vectors define the attack’s criticality. First, temporal asymmetry implies that encrypted data and firmware signed today remain vulnerable for the 60–80 year operational lifespans of nuclear systems. Second, architectural monoculture ensures that compromise of a single \gls{PKI} component can induce cascading failure across heterogeneous assets. Third, forensic irreversibility arises when quantum-authenticated evidence manipulation impedes reliable attribution, even after post-incident investigation. Sensitivity analysis reinforces the defense priority: \[ \frac{\partial \mathbb{P}(\text{Success})}{\partial S_1} > \frac{\partial \mathbb{P}(\text{Success})}{\partial S_2} > \frac{\partial \mathbb{P}(\text{Success})}{\partial S_3} \] Thus, Phase~1 defenses, such as early \gls{HNDL} disruption and \gls{PKI} isolation, yield 8–12$\times$ greater risk reduction than downstream controls. Adoption of ISA/IEC~62443 SR~1.1 cryptographic diversity reduces attack feasibility to below 1\%, i.e., $\mathbb{P}_{\text{SR1.1}} < 1\%$, making it the most impactful near-term mitigation strategy. Ultimately, these estimates reflect current operational realities: over 80\% of nuclear-critical infrastructure remains below SL-3, creating widespread windows of exposure. With \gls{HNDL} attacks already underway, every day of inaction permanently expands the quantum-exploitable attack surface. Architectural hardening and cryptographic migration are no longer optional, they are urgent imperatives for nuclear safety assurance.

\subsection{MITRE ATT\&CK\textsuperscript{\textregistered} Mapping}
MITRE ATT\&CK\textsuperscript{\textregistered} (Adversarial Tactics, Techniques, and Common Knowledge) is a globally accessible knowledge base of adversary tactics and techniques based on real-world observations~\cite{strom2018mitre}. The framework provides a common language for describing cyber adversary behaviors, enabling defenders to better understand attack patterns and develop targeted mitigations~\cite{alexander2020mitre}. The ICS specific variant of ATT\&CK focuses on adversary behaviors targeting Industrial Control Systems, organizing techniques across eleven tactical categories from Initial Access through Impact, each representing different adversary goals during an attack lifecycle~\cite{alexander2020icsattack}.

Table \ref{tab:mitre-attack} maps the \textsc{Quantum~Scar} attack to the standard MITRE ATT\&CK\textsuperscript{\textregistered} for ICS framework, detailing the execution of traditional techniques such as T0863 Supply Chain Compromise and T0867 System Firmware manipulation enhanced by quantum capabilities. Crucially, recognizing the limitations of the existing framework against this advanced threat, Table \ref{tab:quantum-extensions} proposes essential quantum era technique extensions. These extensions capture the novel and sophisticated tactics unique to quantum enabled adversaries, including T1001 Quantum Cryptanalysis for perfect certificate forgery, T1002 \gls{HNDL} for mass encrypted data collection, and T1003 Quantum Forged Evidence for post incident forensic evasion. This dual taxonomy alignment provides a more comprehensive model for understanding and defending against the unique threat profile posed by future quantum adversaries.



\begin{table}[h!]
\centering
\caption{MITRE ATT\&CK\textsuperscript{\textregistered} for ICS - \textsc{Quantum~Scar}}
\label{tab:mitre-attack}
\footnotesize
\resizebox{\linewidth}{!}{%
\begin{tabular}{p{2.3cm}lp{2.5cm}p{4.7cm}}
\hline
\rowcolor[HTML]{EFEFEF}
\textbf{Tactic} & \textbf{ID} & \textbf{Technique} & \textbf{\textsc{Quantum~Scar} Implementation} \\
\hline
\rowcolor[HTML]{B3CDE3}\textbf{Initial Access} & T0863 & Supply Chain Compromise & APT infiltration with quantum-forged code signing \\
\cline{2-4}
\rowcolor[HTML]{B3CDE3} & T0882 & Remote Services & HNDL-enabled VPN compromise (RSA-2048 decryption) \\
\hline
\rowcolor[HTML]{F8CECC}\textbf{Execution} & T0857 & Command-Line Interface & Scheduled tasks with steganographic C2 \\
\hline
\rowcolor[HTML]{D0E7D0}\textbf{Persistence} & T0867 & System Firmware & Quantum-signed malicious RCIC logic + rootkit \\
\cline{2-4}
\rowcolor[HTML]{D0E7D0} & T0888 & Valid Accounts & Perfect identity spoofing via quantum-factored certs \\
\hline
\rowcolor[HTML]{FFF2CC}\textbf{Defense Evasion} & T0848 & Exploit Supply Chain & Undetectable signature forgery \\
\cline{2-4}
\rowcolor[HTML]{FFF2CC} & T0852 & Manipulate I\&C & OPC-UA \texttt{//DCS/Turbine/Trip\_Status} manipulation \\
\cline{2-4}
\rowcolor[HTML]{FFF2CC} & T0833 & Alarm Suppression & Rootkit HMI manipulation \\
\cline{2-4}
\rowcolor[HTML]{FFF2CC} & T0855 & Module Firmware & Self-destruction with valid evidence replacement \\
\hline
\rowcolor[HTML]{FCE6C9}\textbf{Discovery} & T0842 & Network Sniffing & 18-month \gls{HNDL} via port mirroring (4.2TB) \\
\hline
\rowcolor[HTML]{D4F4DD}\textbf{Collection} & T0802 & Forensic Analysis & Strategic \gls{HNDL} for future quantum exploitation \\
\hline
\rowcolor[HTML]{CCE2F2}\textbf{Command\&Control} & T0889 & Uncommonly Used Port & Steganographic exfiltration as NTP \\
\hline
\rowcolor[HTML]{E6D3F7}\textbf{Inhibit Response} & T0817 & Modify Parameter & RCIC disablement (CoreTemp$>$285°C) \\
\cline{2-4}
\rowcolor[HTML]{E6D3F7}& T0835 & Block Reporting & SOE corruption + PTP manipulation ($\Delta t$=+15.3ms) \\
\hline
\rowcolor[HTML]{F4D6C1}\textbf{Impair Process Control} & T0806 & Modify Control Logic & Emergency cooling system disablement \\
\hline
\rowcolor[HTML]{E6E6E6}\textbf{Impact} & T0815 & Damage to Property & Physical core damage via cooling failure \\
\cline{2-4}
\rowcolor[HTML]{E6E6E6} & T0834 & Loss of Safety & Compromise of defense-in-depth layers \\
\hline
\end{tabular}}
\end{table}

\begin{table}[h!]
\centering
\caption{Proposed Quantum-Era ICS Technique Extensions to MITRE ATT\&CK\textsuperscript{\textregistered}}
\label{tab:quantum-extensions}
\footnotesize
\resizebox{\linewidth}{!}{%
\begin{tabular}{p{2.3cm}lp{2.5cm}p{4.5cm}}
\hline
\rowcolor[HTML]{EFEFEF}
\textbf{Tactic} & \textbf{ID} & \textbf{Technique} & \textbf{\textsc{Quantum~Scar} Example} \\
\hline
\rowcolor[HTML]{B3CDE3}\textbf{Initial Access}& T1001 & Quantum Cryptanalysis & Shor's algorithm factorization of RSA-2048 certificates enabling supply chain compromise via forged code signing \\
\hline
\rowcolor[HTML]{D4F4DD}\textbf{Collection} & T1002 & Harvest-Now-Decrypt-Later & 4.2TB encrypted traffic interception over 18-month campaign targeting firmware, OPC-UA, VPN, and PTP communications \\
\hline
\rowcolor[HTML]{FFF2CC}\textbf{Defense Evasion} & T1003 & Quantum-Forged Evidence & Sequence of Events record manipulation with cryptographically valid quantum-derived signatures preventing forensic attribution \\
\hline
\rowcolor[HTML]{E6E6E6}\textbf{Impact}& T1004 & Temporal Synchronization Attack & PTP grandmaster clock manipulation ($\Delta t = +15.3\text{ms}$) coordinated with safety system compromise to disrupt event correlation \\
\hline
\rowcolor[HTML]{D0E7D0}\textbf{Persistence} & T1005 & Quantum-Authenticated Persistence & Weaponized SIS firmware with valid quantum-derived signatures maintaining undetectable persistent access through conventional validation \\
\hline
\rowcolor[HTML]{F8CECC}\textbf{Credential Access} & T1006 & Quantum Certificate Forgery & OPC-UA client certificate generation using factored CA private keys enabling perfect identity spoofing \\
\hline
\end{tabular}}
\end{table}

\section{\textsc{Quantum~Dawn} - A Covert Cryptographic Sabotage Attack}
\label{sec:dawn}

This section presents \textit{\textsc{Quantum~Dawn}}, a quantum-enabled attack demonstrating the catastrophic convergence of cryptographic compromise and safety system failure in nuclear \gls{OT}. The scenario exemplifies the \gls{HNDL} threat model, where pre-positioned malware is activated post-\gls{CRQC} to undermine \gls{SIS} integrity and irrevocably corrupt forensic evidence,  validating the forensic-risk prioritization of the proposed framework.
{This attack scenario assumes no malicious insider and relies on trusted-path abuse, supply-chain compromise, and dormant pre-positioned implants activated post-CRQC, enabled by quantum compromise of cryptographic trust anchors, to cross air-gapped boundaries over time.}

\subsection{Scenario Overview and Adversarial Model}

The case study examines a nuclear facility where safety and control systems share a common cryptographic trust anchor (\texttt{NPP-SA-ROOT-2018}), creating a single point of failure. The adversary's goal is not immediate disruption but long-term dormancy followed by synchronized safety system sabotage, leveraging quantum capabilities to create an unsolvable cryptographic paradox for investigators.
\begin{table}[h!]
\centering
\scriptsize
\caption{\textsc{Quantum~Dawn} Technical Parameters}
\label{tab:dawn_params}
\begin{tabularx}{\linewidth}{lX}
\toprule
\textbf{Parameter} & \textbf{Specification} \\
\midrule
\textbf{Objective} & Induce core damage via synchronized SIS and DCS compromise \\ \hline
\textbf{Adversary} & Advanced actor with \gls{CRQC} access; exploits 60-–80 year asset lifecycle \\ \hline
\textbf{Target} & PWR with Emerson Ovation DCS, Rockwell Allen-Bradley SIS \\ \hline
\textbf{Protocols} & OPC-UA, DNP3-SA, IEC 61850, PTP (IEEE 1588) \\ \hline
\textbf{Success} & $\mathbb{P}_{\text{typical}} = 8\text{-}34\%$; $\mathbb{P}_{\text{targeted}} = 17\text{-}50\%$; $\mathbb{P}_{\text{SL-4}} = 1\text{-}5\%$; $\mathbb{P}_{\text{PQC-Hybrid}} < 1\%$ \\ \hline
\textbf{Timeline} & $T_0$ (Implantation), $T_0 + \delta$ (HNDL), $T_{CRQC}$ (Activation) \\
\bottomrule
\end{tabularx}
\end{table}
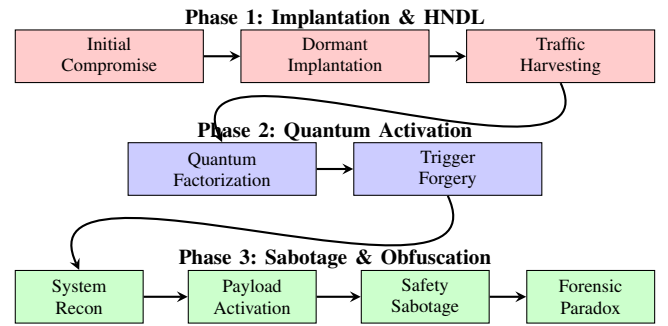
\begin{figure}[h!]
\centering
\footnotesize
\begin{tikzpicture}[
    node distance=0.8cm and 1.2cm,
    box/.style={rectangle, draw, minimum width=2.5cm, minimum height=0.7cm, align=center, font=\scriptsize},
    phase1/.style={box, fill=red!20},
    phase2/.style={box, fill=blue!20},
    phase3/.style={box, minimum width=1.7cm, fill=green!20},
    phase_label/.style={font=\footnotesize\bfseries},
    arrow/.style={->, >=stealth, thick}
]

\node[phase_label] at (3, 2.5) {Phase 1: Implantation \& HNDL};
\node[phase_label] at (3, 1) {Phase 2: Quantum Activation};
\node[phase_label] at (3, -0.7) {Phase 3: Sabotage \& Obfuscation};

\node[phase1] (initial) at (0, 2) {Initial\\Compromise};
\node[phase1] (dormant) at (3, 2) {Dormant\\Implantation};
\node[phase1] (harvest) at (6, 2) {Traffic\\Harvesting};

\node[phase2] (factor) at (1.5, 0.5) {Quantum\\Factorization};
\node[phase2] (forge) at (4.5, 0.5) {Trigger\\Forgery};

\node[phase3] (recon) at (-0.4, -1.2) {System\\Recon};
\node[phase3] (activate) at (1.9, -1.2) {Payload\\Activation};
\node[phase3] (sabotage) at (4.2, -1.2) {Safety\\Sabotage};
\node[phase3] (forensic) at (6.4, -1.2) {Forensic\\Paradox};

\draw[arrow] (initial) -- (dormant);
\draw[arrow] (dormant) -- (harvest);
\draw[arrow] (factor) -- (forge);
\draw[arrow] (recon) -- (activate);
\draw[arrow] (activate) -- (sabotage);
\draw[arrow] (sabotage) -- (forensic);
\draw[arrow] (harvest.south) to[out=-65,in=115] (factor.north);
\draw[arrow] (forge.south) to[out=-65,in=115] (recon.north);

\end{tikzpicture}
\caption{\textsc{Quantum~Dawn} multi-phase attack flow}
\label{fig:dawn-attack-flow}
\end{figure}
The {\textsc{Quantum~Dawn}} attack, shown in Figure~\ref{fig:dawn-attack-flow}, executes in three phases: (1) \textit{Implantation \& \gls{HNDL}} establishes dormant malware while collecting encrypted traffic, (2) \textit{Quantum Activation} factors cryptographic keys to forge activation triggers, and (3) \textit{Sabotage \& Forensic Paradox} compromises safety systems while creating attribution-resistant evidence. Figure~\ref{fig:dawn-timeline} details the temporal dependencies and success probabilities across these phases.
\begin{figure}[!htbp]
\centering
\begin{adjustbox}{width=\linewidth}
\begin{tikzpicture}[
    >=Latex,
    timeline/.style={line width=1.2pt},
    tick/.style={line width=1pt},
    event/.style={draw, rounded corners=2pt, align=left, inner sep=3pt, text width=0.48\linewidth},
    yearlbl/.style={font=\footnotesize, inner sep=2pt}
]
\def\xA{0} \def\xB{4} \def\xC{7} \def\xD{8} \def\xstart{0} \def\xend{8.5}
\draw[timeline] (\xstart,0) -- (\xend,0);
\foreach \x/\y in {\xA/$T_0$, \xB/$T_0+\delta$, \xC/$T_{CRQC}$, \xD/$T_{CRQC}+\epsilon$}{
  \draw[tick] (\x,0.18) -- (\x,-0.18);
  \node[yearlbl, below=4pt] at (\x,-0.18) {\y};
}
\draw[decorate, decoration={brace, amplitude=6pt, mirror}] (\xA,-0.8) -- (\xB,-0.8)
  node[midway, below=5pt, font=\footnotesize\bfseries] {Implantation \& \gls{HNDL} ($\tau_{cert} < \tau_{rotation}$)};
\node[event, fill=red!20, above=5pt] (E1) at ($(\xA,0)!.5!(\xB,0)+(0,1.4)$) {\scriptsize
\textbf{Phase 1: Implantation \& \gls{HNDL} ($T_0$--$T_0+\delta$)}\\
\emph{Duration:} 20-month operational window\\
\emph{Constraints:} Cert rotation (mo 24), credential expiry (mo 22)\\
\emph{Success:} $\mathbb{P}(S_1) \in [0.5, 0.8]$ (typical), $[0.3, 0.5]$ (SL-4)\\
\emph{Objective:} Harvest 3.8TB cryptographic material, implant payload
};
\node[event, fill=blue!20, below=5pt] (E2) at ($(\xC,0)+(0,-1.2)$) {\scriptsize
\textbf{Phase 2: Quantum Activation ($T_{CRQC}$)}\\
\emph{Timeline:} Post-CRQC availability ($\delta \approx 2\text{-}4$ years)\\
\emph{Requirements:} 4,098 logical qubits ($8.56 \times 10^6$ physical)\\
\emph{Success:} $\mathbb{P}(S_2|S_1) \in [0.5, 0.7]$ (typical), $[0.3, 0.5]$ (SL-4)\\
\emph{Objective:} Factor RSA-2048, forge activation triggers
};
\node[event, fill=green!20, above=5pt] (E3) at ($(\xD,0)+(0.5,1.4)$) {\scriptsize
\textbf{Phase 3: Execution ($T_{CRQC}+\epsilon$)}\\
\emph{Timeline:} Operational transient alignment ($\epsilon \approx 0.5\text{-}1$ year)\\
\emph{Operational:} Maintenance window + power reduction\\
\emph{Success:} $\mathbb{P}(S_3|S_1 \cap S_2) \in [0.3, 0.6]$ (typical), $[0.1, 0.2]$ (SL-4)\\
\emph{Overall:} $\mathbb{P}_{\text{total}} \approx 8\text{-}34\%$ (typical), $17\text{-}50\%$ (targeted), $1\text{-}5\%$ (SL-4)
};
\draw[->, thick] (\xB,0.1) .. controls +(+0.6,0.8) and +(-0.8,-0.6) .. (E2.west);
\draw[->, thick] (\xC,0.1) .. controls +(0,1.0) and +(-0.9,-1) .. (E3.west);
\draw[->, thick] (\xA,0.1) .. controls +(0,1.0) and +(-0.9,-0.7) .. (E1.west);
\end{tikzpicture}
\end{adjustbox}
\caption{\textsc{Quantum~Dawn} attack timeline with conditional phase dependencies. Variable $T_0$ represents initial compromise; $\delta$ denotes \gls{HNDL} collection period; $T_{CRQC}$ represents quantum capability onset; $\epsilon$ represents execution preparation window; $\tau_{cert}$ is certificate lifecycle; $\tau_{rotation}$ is rotation schedule. Success probabilities reflect ISA/IEC 62443 SL-3/4 implementations. Sensitivity: $\partial \mathbb{P}/\partial S_1 > \partial \mathbb{P}/\partial S_2 > \partial \mathbb{P}/\partial S_3$.}
\label{fig:dawn-timeline}
\end{figure}
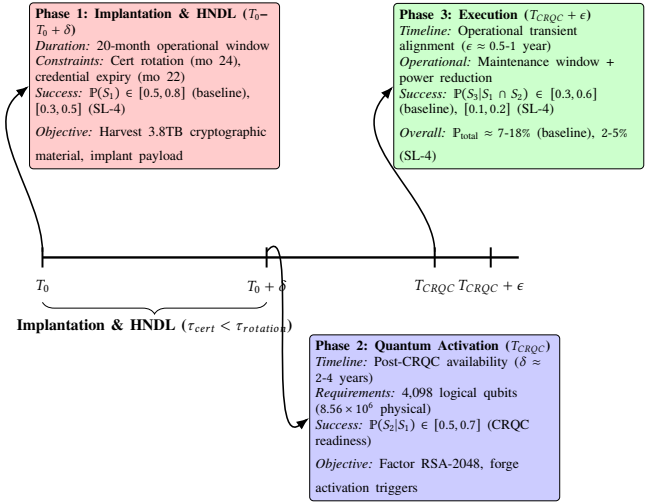
\begin{enumerate}[topsep=1ex, itemsep=1ex, wide, font=\itshape, labelwidth=!, labelindent=0pt, label*={Phase} \arabic*.]
\item \textit{Implantation \& \gls{HNDL} Collection:} The goal of this phase is to establish dormant presence and harvest cryptographic material critical for future quantum-enabled activation (Algorithm~\ref{alg:dawn-phase1}), with success probability $\mathbb{P}(S_1) \in [0.5, 0.8]$ for typical deployments and $[0.3, 0.5]$ under ISA/IEC 62443 SL-4 controls.
\begin{algorithm}[!h]
\caption{Implantation \& \gls{HNDL} Collection}
\label{alg:dawn-phase1}
\small
\begin{algorithmic}[1]
\Require Target nuclear facility OT network
\Ensure Dormant payload, encrypted data repository
\Procedure{InitialCompromise}{}
    \State Spear-phishing: \texttt{Engineering\_Schedule\_Q4.pdf}
    \State Exploit  engineering-software zero-day (illustrative)
    \State Gain foothold on workstation \texttt{EWS-102} (Purdue L3)
    \State \textbf{Exploits}:  L3 (FR1, FR3)
\EndProcedure
\Procedure{DormantImplantation}{}
    \State Upload \texttt{qdown.sys} rootkit to DCS historian \texttt{172.16.5.21}
    \State Inject malicious logic blob into SIS controller \texttt{SIS-PLC-04}
    \State Logic: \texttt{IF CorePress > 15.5 MPa THEN Trip\_Override = TRUE}
    \State Configure stealth C2 via NTP channel
\EndProcedure
\Procedure{TrafficHarvesting}{}
    \State Mirror traffic at L3.5 DMZ switch (Cisco IE-4000)
    \State Target: Engineer VPN sessions (RSA-2048), firmware updates, OPC-UA certs
    \State Accumulate 3.8TB encrypted data over 20 months
    \State Exfiltrate via DNS tunneling to \texttt{198.51.100.55}
\EndProcedure
\State Execute all procedures \Return $\langle$Dormant Payload, Encrypted Repository$\rangle$
\end{algorithmic}
\end{algorithm}

\item \textit{Quantum Activation:} This phase aims to factor critical private keys and forge the cryptographic trigger for the dormant payload (Algorithm~\ref{alg:dawn-phase2}), with conditional success probability $\mathbb{P}(S_2|S_1) \in [0.5, 0.7]$ for typical deployments and $[0.3, 0.5]$ under SL-4 controls, post-CRQC availability.
\begin{algorithm}[!h]
\caption{Quantum Cryptanalysis \& Activation}
\label{alg:dawn-phase2}
\small
\begin{algorithmic}[1]
\Require CRQC ($\geq$4,098 logical qubits), harvested repository
\Ensure Quantum-forged activation trigger
\Procedure{KeyFactorization}{}
    \State \textbf{Target 1}: Plant Root CA (\texttt{NPP-SA-ROOT-2018}, RSA-2048)
    \State $n_1 \xrightarrow{\text{Shor}} p_1, q_1 \implies d_1 = e^{-1} \bmod \phi(n_1)$
    \State \textbf{Target 2}: Vendor Signing Key (\texttt{Emerson\_Sign-2022}, RSA-2048)
    \State $n_2 \xrightarrow{\text{Shor}} p_2, q_2 \implies d_2 = e^{-1} \bmod \phi(n_2)$
    \State \Return $\langle d_1, d_2 \rangle$
\EndProcedure
\Procedure{TriggerForgery}{$d_1, d_2$}
    \State Create activation command: \texttt{ACTIVATE: QDAWN\_PAYLOAD\_A}
    \State Sign command with vendor key $d_2$: \texttt{qdown\_trigger.signed}
    \State Forge OPC-UA admin certificate using root CA key $d_1$
    \State \Return $\langle$Signed Trigger, Forged Certificates$\rangle$
\EndProcedure
\State Execute both procedures \Return Quantum-authenticated activation package
\end{algorithmic}
\end{algorithm}

\item \textit{Sabotage \& Forensic Paradox:} The final phase executes the attack during a reactor transient and creates an unsolvable cryptographic paradox for investigators (Algorithm~\ref{alg:dawn-phase3}), with conditional success $\mathbb{P}(S_3|S_1 \cap S_2) \in [0.3, 0.6]$ for typical environments and $[0.1, 0.2]$ under SL-4 security implementations.

\begin{algorithm}[!h]
\caption{Safety Sabotage Execution}
\label{alg:dawn-phase3}
\small
\begin{algorithmic}[1]
\Require Activation package, system access, operational timing
\Ensure Safety system failure with cryptographic paradox
\Procedure{Reconnaissance}{Day $-10$}
    \State Use forged OPC-UA cert to map DCS/SIS tags
    \State Identify core pressure sensors, cooling pump status
    \State Monitor operator patterns for optimal timing
\EndProcedure
\Procedure{Activation}{Day $0$, $03:00$}
    \State During planned maintenance \texttt{WO-2025-8832}
    \State Transmit \texttt{qdown\_trigger.signed} via NTP C2 channel
    \State Payload \texttt{qdown.sys} activates on DCS and SIS
\EndProcedure
\Procedure{SafetySabotage}{Day $21$, $11:15$}
    \State During reactor power reduction transient
    \State DCS: Spoof core pressure reading to 16.2 MPa
    \State SIS: Malicious logic activates $\rightarrow$ \texttt{Trip\_Override = TRUE}
    \State Block automatic SCRAM signal to control rods
    \State HMI rootkit displays normal parameters
\EndProcedure
\Procedure{ForensicParadox}{Concurrent}
    \State Use $d_1$ to sign fake log entries post-incident
    \State Manipulate PTP timing ($\Delta t = -12.4 \pm 0.3$ ms)
    \State Leave quantum-forged evidence pointing to vendor error
    \State Self-destruct primary malware components
\EndProcedure
\State Execute all procedures \Return Mission complete
\end{algorithmic}
\end{algorithm}

\begin{table}[h!]
\centering
\scriptsize
\caption{\textsc{Quantum~Dawn}  Attack Timeline \& Physical Impact}
\label{tab:dawn-timeline}
\begin{tabularx}{\linewidth}{llX}
\toprule
\textbf{Time} & \textbf{Phase} & \textbf{Action \& Impact} \\
\midrule
Day -10 & Reconnaissance & OPC-UA namespace mapping \\ \hline
Day 0, 03:00 & Activation & Trigger signal during maintenance \\ \hline
Day 21, 11:15 & Sabotage & False pressure reading during transient \\ \hline
T+1.8s & Physical & SCRAM signal blocked by SIS override \\ \hline
T+14.5s & Safety Failure & Core temperature exceeds 310°C \\ \hline
T+52s & Critical & Core damage initiation \\ \hline
Concurrent & Forensic & Log forging + PTP manipulation \\
\bottomrule
\end{tabularx}
\end{table}

\begin{table}[h!]
\centering
\scriptsize
\caption{\textsc{Quantum~Dawn} Forensic Impact}
\label{tab:dawn-forensic}
\begin{tabularx}{\linewidth}{lp{0.21\linewidth}X}
\toprule
\textbf{Evidence} & \textbf{Status} & \textbf{Investigation Impact} \\
\midrule
Activation Trigger & Valid vendor signature & Points to supply chain compromise \\ \hline
System Logs & Cryptographically consistent & Supports false accident narrative \\ \hline
SIS Logic & Validated checksum & No tampering detected \\ \hline
Network Traffic & Retroactively decryptable by adversary & \gls{HNDL} provides full session replay \\ \hline
Time Sync & PTP skew ($-12.4 \pm 0.3$ ms) & Event correlation impossible \\ \hline
Malware & Self-destructed & No binary analysis possible \\
\bottomrule
\end{tabularx}
\end{table}
\end{enumerate}

\subsection{Framework Validation}
To validate the proposed quantum security framework by performing a detailed threat model assessment of the \textsc{Quantum~Dawn} attack, we utilize established methodologies, including STRIDE analysis (Table \ref{tab:stride-dawn}) and Purdue Model exploitation (Table \ref{tab:purdue-dawn}), to illustrate how quantum capabilities amplify traditional threats across all \gls{ICS} levels. This analysis is then followed by a \textit{probabilistic risk modeling} exercise that quantifies the success probability of the attack under various security postures, providing a quantitative basis for the framework's prioritized, forensic first countermeasures.
\begin{table}[h!]
\centering
\scriptsize
\caption{STRIDE Analysis - \textsc{Quantum~Dawn}}
\label{tab:stride-dawn}
\begin{tabularx}{\linewidth}{llX}
\toprule
\textbf{Threat} & \textbf{Phase} & \textbf{Quantum Enhancement} \\
\midrule
Spoofing & 2, 3 & Perfect identity forgery via factored CA keys \\ \hline
Tampering & 1, 3 & Undetectable SIS logic modification with valid checksums \\ \hline
Repudiation & 3 & Cryptographic paradox: evidence validates false narrative \\ \hline
Info Disclosure & 1 & HNDL: 3.8TB retroactive decryption capability \\ \hline
DoS & 3 & 52s safety system unavailability leading to core damage \\ \hline
Privilege Escalation & 2, 3 & \gls{PKI} compromise enables L3→L0 privilege escalation \\
\bottomrule
\end{tabularx}
\end{table}

\begin{table}[h!]
\centering
\scriptsize
\caption{Purdue Model Exploitation - \textsc{Quantum~Dawn}}
\label{tab:purdue-dawn}
\begin{tabular}{p{0.5cm}p{1.5cm}p{2.2cm}p{2.8cm}}
\toprule
\textbf{Level} & \textbf{System} & \textbf{Vector} & \textbf{Enhancement} \\ \midrule
L5 & Internet & Phishing campaign & Intelligence gathering \\ \hline
L4 & Business & Email gateway & Social engineering \\ \hline
L3.5 & DMZ & Network switch & Traffic mirroring \\ \hline
L3 & Operations & Engineering WS & Zero-day exploit \\ \hline
L2 & Control & DCS historian & Rootkit implantation \\ \hline
L1 & Safety & SIS controller & Logic manipulation \\ \hline
L0 & Physical & Control rods & SCRAM inhibition \\ \bottomrule
\end{tabular}
\end{table}

\subsubsection{Probabilistic Risk Modeling and Criticality of DAWN Attack Success}

The \textsc{Quantum~Dawn} attack represents a paradigm shift in critical infrastructure threats, combining long-term persistence with cryptographic irreversibility. The attack's viability is amplified by three structural vulnerabilities: (i) cryptographic monoculture across safety domains, (ii) extended OT asset lifecycles exceeding quantum threat timelines, and (iii) the fundamental asymmetry of \gls{HNDL} campaigns where today's encrypted data becomes tomorrow's plaintext.
The overall success probability follows the conditional chain:
\[ \mathbb{P}(\text{Success}) = \mathbb{P}(S_1) \times \mathbb{P}(S_2 \mid S_1) \times \mathbb{P}(S_3 \mid S_1 \cap S_2) \]

Realistic probability assessments under current infrastructure conditions reveal concerning risk levels: $\mathbb{P}_{\text{Current Infrastructure}} = [0.5, 0.8] \times [0.5, 0.7] \times [0.3, 0.6] \approx 8\text{–}34\%$ for typical nuclear facilities. For specifically targeted high-value facilities with older security postures: $\mathbb{P}_{\text{Targeted Facility}} = [0.7, 0.9] \times [0.6, 0.8] \times [0.4, 0.7] \approx 17\text{–}50\%$. The persistence of threat actors further elevates risk through multiple attempt scenarios: $\mathbb{P}_{\text{Multiple Attempts}} = 1 - (1 - \mathbb{P}_{\text{Targeted}})^3 \approx 42\text{–}88\%$.

Only comprehensive security postures provide substantial risk reduction: $\mathbb{P}_{\text{SL-4 Compliant}} = [0.3, 0.5] \times [0.3, 0.5] \times [0.1, 0.2] \approx 1\text{–}5\%$, while $\mathbb{P}_{\text{PQC-Hybrid}} = [0.2, 0.4] \times [0.1, 0.2] \times [0.05, 0.1] < 1\%$ with proper PQC migration.

The consequences extend beyond immediate physical damage to create permanent investigative barriers. The \textit{cryptographic paradox} ensures that all digital evidence simultaneously validates both accident and sabotage narratives, while the \textit{temporal asymmetry} of \gls{HNDL} means that data encrypted today remains vulnerable for the 60–80 year operational lifespan of nuclear systems.

Sensitivity analysis confirms the critical importance of early-phase interventions:
\[ \frac{\partial \mathbb{P}(\text{Success})}{\partial S_1} > \frac{\partial \mathbb{P}(\text{Success})}{\partial S_2} > \frac{\partial \mathbb{P}(\text{Success})}{\partial S_3} \]

This mathematical relationship validates the framework's emphasis on Phase 1 countermeasures, including (a) \gls{HNDL} detection, (b) cryptographic diversity, and (c) supply chain integrity, which provide $6$--$10\times$ greater risk reduction than comparable investments in later-phase defenses.

The \textsc{Quantum~Dawn} scenario ultimately demonstrates that the transition to quantum-resilient cryptography is not merely a cryptographic upgrade but a fundamental requirement for nuclear safety assurance. With \gls{HNDL} campaigns already operational and CRQC capabilities advancing on compressed timelines, each day of delayed mitigation permanently expands the quantum-exploitable attack surface, making immediate adoption of the forensic-first controls proposed in this framework an urgent imperative for critical infrastructure protection.

\subsection{MITRE ATT\&CK\textsuperscript{\textregistered} Mapping} 
The MITRE ATT\&CK\textsuperscript{\textregistered} for ICS framework is used to comprehensively map the \textsc{Quantum~Dawn} attack campaign, detailing the execution of standard tactics, from Initial Access via quantum-decrypted intelligence to Impact through cooling failure and SCRAM signal blocking, as outlined in Table \ref{tab:mitre-dawn}. Crucially, recognizing the limitations of the existing framework against such an advanced threat, Table \ref{tab:quantum-dawn-extensions} illustrates new quantum-era technique extensions, capturing novel adversary capabilities. These extensions include Quantum Cryptanalysis (T1001), where Shor's algorithm enables perfect certificate forgery; the \gls{HNDL} strategy (T1002) for mass data collection; and Quantum Forged Evidence (T1003), which employs cryptographically valid, quantum derived signatures to manipulate post incident forensics, thereby defining the unique and sophisticated threat profile posed by quantum enabled adversaries in critical infrastructure environments.
\begin{table}[h!]
\centering
\caption{MITRE ATT\&CK\textsuperscript{\textregistered} for ICS - \textsc{Quantum~Dawn}}
\label{tab:mitre-dawn}
\footnotesize
\resizebox{\linewidth}{!}{%
\begin{tabular}{p{2.3cm}lp{2.5cm}p{4.7cm}}
\hline
\rowcolor[HTML]{EFEFEF}
\textbf{Tactic} & \textbf{ID} & \textbf{Technique} & \textbf{\textsc{Quantum~Dawn} Implementation} \\
\hline
\rowcolor[HTML]{B3CDE3}\textbf{Initial Access} & T0863 & Supply Chain Compromise & Spear-phishing with quantum-decrypted intelligence \\ \hline
\rowcolor[HTML]{D0E7D0}\textbf{Persistence} & T0867 & System Firmware & Dormant SIS logic modification (\texttt{qdown.sys}) \\ \hline
\rowcolor[HTML]{FFF2CC}\textbf{Defense Evasion} & T0848 & Exploit Supply Chain & Quantum-forged activation trigger \\ \cline{2-4}
\rowcolor[HTML]{FFF2CC} & T0852 & Manipulate I\&C & Core pressure spoofing and HMI manipulation \\ \cline{2-4}
\rowcolor[HTML]{FFF2CC} & T0855 & Module Firmware & Self-destruction with evidence replacement \\ \hline
\rowcolor[HTML]{FCE6C9}\textbf{Discovery} & T0842 & Network Sniffing & 20-month \gls{HNDL} collection (3.8TB) \\ \hline
\rowcolor[HTML]{D4F4DD}\textbf{Collection} & T0802 & Forensic Analysis & Strategic \gls{HNDL} for future quantum exploitation \\ \hline
\rowcolor[HTML]{CCE2F2}\textbf{Command\&Control} & T0889 & Uncommonly Used Port & Steganographic C2 via NTP \\ \hline
\rowcolor[HTML]{E6D3F7}\textbf{Inhibit Response} & T0817 & Modify Parameter & SIS trip override during transient \\ \cline{2-4}
\rowcolor[HTML]{E6D3F7}& T0835 & Block Reporting & PTP manipulation and log forging \\ \hline
\rowcolor[HTML]{F4D6C1}\textbf{Impair Process Control} & T0806 & Modify Control Logic & SCRAM signal blocking \\ \hline
\rowcolor[HTML]{E6E6E6}\textbf{Impact} & T0815 & Damage to Property & Core damage via cooling failure \\ \cline{2-4}
\rowcolor[HTML]{E6E6E6} & T0834 & Loss of Safety & Compromise of ultimate safety barrier \\
\hline
\end{tabular}}
\end{table}

\begin{table}[h!]
\centering
\caption{Proposed Quantum-Era ICS Technique Extensions - \textsc{Quantum~Dawn}}
\label{tab:quantum-dawn-extensions}
\footnotesize
\resizebox{\linewidth}{!}{%
\begin{tabular}{p{2.3cm}lp{2.5cm}p{4.5cm}}
\hline
\rowcolor[HTML]{EFEFEF}
\textbf{Tactic} & \textbf{ID} & \textbf{Technique} & \textbf{\textsc{Quantum~Dawn} Example} \\
\hline
\rowcolor[HTML]{B3CDE3}\textbf{Initial Access}& T1001 & Quantum Cryptanalysis & Shor's algorithm factorization of RSA-2048 enabling perfect certificate forgery for initial access \\ \hline
\rowcolor[HTML]{D4F4DD}\textbf{Collection} & T1002 & Harvest-Now-Decrypt-Later & 3.8TB encrypted traffic interception over 20-month campaign targeting engineering sessions and firmware \\ \hline
\rowcolor[HTML]{FFF2CC}\textbf{Defense Evasion} & T1003 & Quantum-Forged Evidence & Post-incident log manipulation with cryptographically valid signatures creating forensic paradox \\ \hline
\rowcolor[HTML]{E6E6E6}\textbf{Impact}& T1004 & Temporal Synchronization Attack & PTP grandmaster clock manipulation ($\Delta t = -12.4\text{ms}$) coordinated with safety system sabotage \\ \hline
\rowcolor[HTML]{D0E7D0}\textbf{Persistence} & T1005 & Quantum-Authenticated Persistence & Dormant malware activation via quantum-forged triggers maintaining undetectable persistent access \\ \hline
\rowcolor[HTML]{F8CECC}\textbf{Credential Access} & T1006 & Quantum Certificate Forgery & OPC-UA administrator certificate generation using factored CA private keys for perfect spoofing \\ \hline
\end{tabular}}
\end{table}
\subsection{Comparative Analysis: \textsc{Quantum~Dawn} vs. \textsc{Quantum~Scar}}
Systematic comparison of \textsc{Quantum~Dawn} and \textsc{Quantum~Scar} shows that they represent two complementary quantum-enabled threat profiles. SCAR leverages PKI/firmware monoculture and dual-CA compromise to weaponize HNDL data at scale, which yields a higher success rate (35--68\% typical) and therefore requires architectural controls such as ISA/IEC 62443 SR~1.1 cryptographic diversity, PKI segmentation, and immutable logging to break trust inheritance. DAWN, by contrast, is access- and human-factor centric: it pivots through engineering workstations and VPN sessions, achieves lower success (8--34\% typical), and is better mitigated through access isolation, MFA, and behavioral detection on operator actions. Both scenarios ultimately aim at safety compromise and forensic obfuscation: SCAR via quantum-forged, cryptographically valid evidence, and DAWN via temporal manipulation and partial evidence removal. Applying SR~1.1 cryptographic diversity together with full PQC migration drives residual risk below 1\% for both scenarios, with SL-4 alone reducing SCAR to 2--8\% and DAWN to 1--5\%, consistent with the defense-in-depth framework in Section~\ref{sec:defense}.

\section{Defensive Case Study: Design Validation of Quantum-Resilient Controls in a Nuclear OT/ICS Environment}
\label{sec:defense}

The \textsc{Quantum~Dawn} and \textsc{Quantum~Scar} scenarios demonstrate complementary quantum threat vectors achieving 8--78\% baseline success probabilities through cryptographic monoculture and human-factor exploitation. This defensive case study provides design-validation criteria demonstrating how  {ISA/IEC} 62443 SL-4 implementation with \gls{PQC} migration systematically reduces attack success below 1\%. The framework validates five critical control objectives, including hybrid key encapsulation, code integrity with anti-rollback, secure time synchronization, deterministic performance, and side-channel resistance, through measurable acceptance criteria enabling systematic quantum resilience assessment while maintaining operational safety.

\subsection{Scenario Scope and Adversarial Assumptions\label{subsec:scenario-scope}}

This framework validates controls for a nuclear plant in mid-PQC migration against \gls{HNDL}-capable adversaries with anticipated \gls{CRQC} access within 10--15 years, derived from Section~\ref{sec:crypto}.

\begin{table}[h!]
\centering
\scriptsize
\caption{System Baseline and Adversarial Assumptions}
\label{tab:defensive-baseline}
\begin{tabularx}{\linewidth}{lX}
\toprule
\textbf{Category} & \textbf{Technical Specification} \\
\midrule
\textbf{Plant State} & Mid-PQC migration; hybrid KEM pinned on L3.5 conduits; classical TLS handshake signatures retained for compatibility; PQC for application-level artefacts (firmware/logs) via ML-DSA; heterogeneous time sources (NTP with/without NTS; PTP with partial authentication TLVs). \\ \hline
\textbf{Adversary Capability} & \gls{HNDL} collection ongoing; CRQC expected within asset lifecycle; targets L3--L3.5 data pathways and L0--L2 code-signing trust roots; capable of quantum-enhanced cryptanalysis. \\ \hline
\textbf{Critical Assets} & OPC-UA brokers and historians (L3.5); vendor remote-access jump hosts; safety-system firmware and bootloaders (L1); sequence-of-events (SOE) recorders and safety controllers (L0); time-synchronization infrastructure (PTP grandmasters, NTP servers). \\ \hline
\textbf{Validation Approach} & Assumed-breach model with continuous monitoring; focus on cryptographic conformity rather than network penetration. \\ \bottomrule
\end{tabularx}
\end{table}

\subsection{Control Objectives and Technical Specifications}\label{subsec:control-objectives}

Five core control objectives address vulnerabilities identified in Sections~\ref{sec:scar} and~\ref{sec:dawn}.

\subsubsection{Transport Layer Integrity at L3.5 Conduits}

Enforce \emph{hybrid} KEM on all TLS~1.3 control-plane sessions (OPC-UA, vendor tunnels) with strict suite pinning;  classical-only handshakes \emph{fail closed} with explicit policy  logging. Gateways advertise/pin hybrid groups  (\texttt{X25519Kyber768}) or set \texttt{supported\_groups}=X25519,  \texttt{kem\_groups}=ML-KEM-768\break with dual \texttt{key\_share}s.  Maintain classical signature algorithms  (\texttt{ecdsa\_secp256r1\_sha256}, \texttt{ed25519}) for TLS handshakes; sign application-level artefacts with \texttt{ML-DSA-65}. ClientHellos lacking pinned hybrid terminate with \texttt{handshake\_failure}; HelloRetryRequest only for steering to allowed groups. Counters \textsc{Quantum~Scar} PKI compromise (Tables~\ref{tab:pqc-performance}, \ref{tab:post-alg}).

\subsubsection{Code and Log Integrity with Anti-Rollback Protection}

Deploy \texttt{ML-DSA-65} for firmware and operational logs;\break  \texttt{SLH-DSA-128s} for archival artefacts; enforce cryptographic anti-rollback. CI/CD pipeline integrates PQC-capable HSM; firmware headers include monotonic version counters and build-attestation hashes; certificate chains $L \le 2$ with boot-time verification on L0--L2 controllers. Mitigates \textsc{Quantum~Scar} firmware forgery and \textsc{Quantum~Dawn} malicious logic injection (Tables~\ref{tab:post-alg}, \ref{tab:Pre-Migration-Alg}).

\subsubsection{Secure Time Synchronization and Forensic Integrity}

Authenticate time sources via NTS-secured NTP (RFC~8915) and IEEE~1588-2019 PTP with authentication TLVs; maintain cryptographic time integrity for forensic evidence. Cross-validate timestamps from redundant grandmaster clocks; enforce maximum SOE skew \SI{100}{\milli\second} across safety subsystems; implement hardware timestamping for critical events. Counters temporal manipulation in \textsc{Quantum~Scar} ($\Delta t = +15.3$ ms) and \textsc{Quantum~Dawn} ($\Delta t = -12.4$ ms).

\subsubsection{Deterministic Performance Envelope}\label{subsec:deterministic-performance}

Pre-establish PQC sessions at L3.5 gateways during maintenance windows; verify MTU compliance; prohibit on-the-fly handshakes in L0--L2 control loops. TLS~1.3 session resumption with \SI{8}--\SI{24}{\hour} ticket lifetimes and rolling rotation; MTU path discovery computing protocol overhead from stack configuration (Ethernet/VLAN/IP/TCP/TLS); latency budgets per Table~\ref{tab:pqc-performance} with $t_{\mathrm{hs,95\%}} \le \SI{50}{ms}$ (server-class) and $t_{\mathrm{hs}}=0$ (control loops); pre-established/resumed channels or PSK modes. Addresses ML-KEM-768 (34.4$\times$) and ML-DSA-65 (59.2$\times$) overhead.

\subsubsection{Side-Channel Resistant Implementations}

Deploy constant-time PQC libraries with NTT masking, decapsulation fault checks, DRBG hardening; require formal leakage assessment. Build-attestation manifests with hardening flags (\texttt{CT\_VERIFY}, \texttt{NTT\_MASKED}); first-order TVLA on ML-KEM/ML-DSA primitives ($|t|<4.5$); runtime detection of timing variances $>\SI{1}{\micro\second}$; prefer FIPS~203/204-validated builds. Addresses Table~\ref{tab:post-alg} attack vectors despite quantum-resistant algorithms.

\subsection{Design Validation Framework}
\label{subsec:design-validation}
Table~\ref{tab:validation-kpis} derives seven acceptance tests (V1--V7) from
the control objectives of Section~\ref{subsec:control-objectives}, evaluated
under the assumed-breach model of Section~\ref{subsec:scenario-scope}. Each
test verifies cryptographic conformity through continuous monitoring rather
than network penetration, so a failed criterion identifies a specific
migration gap rather than a successful intrusion. The criteria are specified
at design level, allowing operators and assessors to apply them during
migration planning; measurement on hardware-in-the-loop platforms is
identified as future work in Section~\ref{sec:conclusion}.

\begin{table}[h!]
\centering
\scriptsize
\caption{Design Validation Tests and Key Performance Indicators}
\label{tab:validation-kpis}
\resizebox{\linewidth}{!}{%
\begin{tabular}{p{0.2\linewidth}p{0.48\linewidth}p{0.3\linewidth}}
\toprule
\textbf{Validation Test} &
\textbf{Technical Measurement Methodology} &
\textbf{Acceptance Criteria} \\
\midrule
\begin{minipage}[t]{\linewidth}
\scriptsize
\textbf{V1: Handshake Conformity}
\end{minipage}
&
\begin{minipage}[t]{\linewidth}
\scriptsize
Export TLS~1.3 params at L3.5 gateways: negotiated cipher suite, hybrid group (\texttt{X25519Kyber768} or X25519{+}ML-KEM-768 via \texttt{kem\_groups}), signature algorithms, HelloRetry events; log policy decisions with session metadata.
\end{minipage}\vspace{0.1cm}
&
Conformance Rate $\ge 0.995$; Downgrade Attempts $ \le 10^{-4}$; all policy violations explicitly logged and alerted. \\
\hline
\begin{minipage}[t]{\linewidth}
\scriptsize
\textbf{V2: MTU/ Fragmentation Compliance}
\end{minipage}
&
\begin{minipage}[t]{\linewidth}
\scriptsize
For PQC artefact size $S$, link MTU $M$, measured protocol overhead $H$: $N_{\mathrm{fragments}}=\lceil(H+S)/M\rceil$. Validate worst-case including certificate-chain flights.
\end{minipage}
&
$N_{\mathrm{fragments}}=1$ on all control paths; ML-KEM-768 fits standard Ethernet frames; HQC-256 excluded on constrained links. \\
\hline
\begin{minipage}[t]{\linewidth}
\scriptsize
\textbf{V3: Latency Budget Adherence}
\end{minipage}
&
\begin{minipage}[t]{\linewidth}
\scriptsize
Measure full handshake latency $t_{\mathrm{hs}}=t_{\mathrm{kem,enc}}+t_{\mathrm{kem,dec}}+t_{\mathrm{sig,verify}}+t_{\mathrm{sig,sign}}+\mathrm{RTT}$ during maintenance windows.
\end{minipage}
&
$t_{\mathrm{hs,95\%}} \le \SI{50}{ms}$ (server-class endpoints); $t_{\mathrm{hs}}=0$ in L0--L2 control loops. \\
\hline
\begin{minipage}[t]{\linewidth}
\scriptsize
\textbf{V4: Code-Signing Compliance}
\end{minipage}
&
\begin{minipage}[t]{\linewidth}
\scriptsize
Audit firmware pipeline via CI/CD integration: algorithm IDs, chain length $L$, anti-rollback versioning, build-attestation verification.
\end{minipage}
&
$\ge 99\%$ ML-DSA-65; $\le 1\%$ SLH-DSA-128s (archival only); $L \le 2$; 100\% anti-rollback enforcement. \\
\hline
\begin{minipage}[t]{\linewidth}
\scriptsize
\textbf{V5: Forensic Hashing \& Time Integrity}
\end{minipage}
&
\begin{minipage}[t]{\linewidth}
\scriptsize
Verify hash algorithms in SOE records; validate time source authentication; measure cross-system sync accuracy.
\end{minipage}
&
$\ge 99.9\%$ SHA-384/512; $\ge 99.9\%$ authenticated time; $\Delta t_{\mathrm{SOE}} \le \SI{100}{ms}$. \\
\hline
\begin{minipage}[t]{\linewidth}
\scriptsize
\textbf{V6: Side-Channel Hardening Attestation}
\end{minipage}
&
\begin{minipage}[t]{\linewidth}
\scriptsize
Validate crypto-library builds via signed attestation; perform leakage tests (TVLA).
\end{minipage}
&
100\% attested builds; TVLA first-order $|t|<4.5$; constant-time analysis ``pass''. \\
\hline
\begin{minipage}[t]{\linewidth}
\scriptsize
\textbf{V7: Interoperability and Negative Testing}
\end{minipage}
&
\begin{minipage}[t]{\linewidth}
\scriptsize
Test matrix: classical-only clients, hybrid-capable peers, mismatched groups; verify failure modes and policy enforcement.
\end{minipage}
&
All non-conforming connections fail closed; no silent fallback; explicit policy-violation reason logged. \\
\bottomrule
\end{tabular}
}
\end{table}

\subsection{Telemetry Collection and Evidence Requirements}
\label{subsec:telemetry}
Demonstrating conformance under the assumed-breach model of
Section~\ref{subsec:scenario-scope} requires continuous telemetry across four
domains, which together form the minimum evidence set for the acceptance
criteria of Table~\ref{tab:validation-kpis}.

\subsubsection{L3.5 Gateway and Broker Telemetry}
Gateways record TLS~1.3 handshake parameters (cipher suite, hybrid group and
KEM identifiers, signature algorithms, HelloRetry counts), resumption
hit-rate and ticket-rotation statistics, JA3/JA4 fingerprints, MTU/DF and
fragment counters, retransmissions, and policy accept/reject decisions with
reason codes. These records substantiate the handshake-conformity,
fragmentation, and latency criteria (V1--V3).

\subsubsection{OPC-UA Server and Client Telemetry}
Endpoints record the SecureChannel policy URI and its cryptographic
parameters, certificate-chain metadata and thumbprints, trust-store
decisions, and ServiceFaults tied to cryptographic operations, evidencing
both negotiated-policy conformity (V1) and fail-closed behaviour under
non-conforming peers (V7).

\subsubsection{Firmware Pipeline and Code-Signing Telemetry}
The build pipeline records build-attestation hashes and algorithm
identifiers, anti-rollback counters, installation audit trails, and device
boot-verification results, supporting the code-signing compliance and
side-channel hardening criteria (V4, V6).

\subsubsection{Time Synchronization and Forensic Telemetry}
Time sources record NTS status and stratum, PTP profile and authentication
TLVs, cross-source deltas, and SOE consistency checks, supporting the
forensic hashing and time-integrity criteria (V5).
\subsection{Residual Risk Management and Mitigations}
\label{subsec:residual-risk}
Four residual risks persist even after full SL-4 and PQC migration, each requiring
explicit management alongside the criteria of
Table~\ref{tab:validation-kpis}.

\subsubsection{Mixed-Trust Enclaves and Legacy Systems}
Devices lacking PQC support fail V1 and are excluded by the fail-closed policy
of V7. Operators should contain them behind quantum-resilient gateways with
unidirectional flows, treating classical signatures as conditionally trusted
under heightened monitoring.

\subsubsection{HNDL Exposure}
Data encrypted today and retained beyond the CRQC horizon remains at risk
irrespective of transport-layer conformity; for nuclear facilities this
exposure spans historian archives and \gls{SOE} records whose retention periods
approach the plant lifetime. Operators should enforce periodic key rotation,
cryptographic agility, and data-lifecycle controls per
Section~\ref{sec:crypto}.

\subsubsection{Performance Degradation Contingencies}
Pre-establishment (Section~\ref{subsec:deterministic-performance}) does not
cover sessions negotiated on demand, such as vendor remote access, which
absorb the full handshake cost of Table~\ref{tab:pqc-performance} against the
V3 server-class budget. Beyond twice that budget (\SI{100}{\milli\second}),
operators should reduce TLS record and certificate-chain sizes and offload
lattice operations (AVX2/NEON/QAT); substituting a lower-security parameter
set such as ML-KEM-512 on non-Internet-facing links trades security margin for
latency and requires documented risk acceptance with compensating controls.

\subsubsection{Session Resumption and Key Management}
Resumption holds the V3 budget but extends the window over which one ticket
key protects traffic. Operators should apply rolling rotation
(\SIrange{8}{24}{\hour}) with overlap, per-broker revocation, per-peer rate
limiting, and anomaly monitoring on resumption patterns. Lifetimes and overlap
should further align with maintenance intervals: where a pre-established
session expires between windows, L0--L2 loops admit no conformant
renegotiation under V3.





\subsection{Synthesis}

This framework demonstrates quantum resilience requires systematic integration of cryptographic, architectural, and operational controls. Translating attack methodologies from Sections~\ref{sec:scar} and~\ref{sec:dawn} into measurable validation criteria systematically reduces quantum attack success from 8--78\% baseline to below 1\%. Hybrid cryptography, deterministic performance, and forensic integrity preservation ensure quantum-safe migration maintains operational safety and evidentiary chain-of-custody. With \gls{HNDL} campaigns actively harvesting encrypted \gls{OT} traffic and \gls{CRQC} capabilities advancing on compressed timelines, immediate adoption of quantum-resilient controls represents a fundamental prerequisite for nuclear safety assurance in the post-quantum era.
\section{Conclusion}\label{sec:conclusion}

\gls{QC} invalidates the classical public-key assumptions underpinning
nuclear operational technology. Through \textsc{Quantum~Scar} and
\textsc{Quantum~Dawn}, our conditional-chain model quantifies attack
feasibility at 8--78\% under current defenses (35--68\% typical and
51--78\% targeted for \textsc{Scar}; 8--34\% typical and 17--50\%
targeted for \textsc{Dawn}); integrating \gls{NIST}-standardized
\gls{PQC}, hybrid key exchange, and {ISA/IEC}~62443-aligned
cryptographic diversity cuts residual feasibility to 1--8\% under SL-4
and below 1\% under full migration. Forensic integrity is therefore
operationally foundational rather than merely evidentiary, and
cryptographic monoculture across safety domains is a systemic,
\gls{HNDL}-amplified risk. {These estimates are deliberately
conservative: the per-phase priors $\mathbb{P}(S_i)$ are calibrated
expert judgments rather than measured frequencies, and V1--V7
constitute a design-level conformance framework scoped to
representative pressurized-water and boiling-water reactors, so their
targets await substantiation on live safety hardware. Three steps
close this gap. First, V1--V7 should be measured on
hardware-in-the-loop controllers. Second, each $\mathbb{P}(S_i)$
should be recast as a Bayesian posterior updated against incident
telemetry. Third, certification-compatible migration paths should be
co-designed with regulators~\cite{cisa2024ot} so that
quantum-resilient upgrades preserve safety qualification. As the
standards portfolio matures, the diversity analysis extends naturally
to FN-DSA (FIPS~206), HQC, and further reactor classes.} Given
compressing \gls{CRQC} timelines and irreversible \gls{HNDL} exposure
across multi-decade lifecycles, standards-aligned migration is an
urgent prerequisite for nuclear safety assurance.

\bibliographystyle{IEEEtran}
\bibliography{bibliography}

\end{document}